%% file: main.tex
\documentclass[whitelogo]{tudelft-report}

\usepackage{natbib}
\usepackage{changes}
\usepackage{pdfpages}
\usepackage{subcaption}
\usepackage{braket}
\usepackage{bm}
\usepackage{tikz,pgfplots,pgfplotstable}
\pgfplotsset{compat=1.9,width=\linewidth, height=8cm}
\usepackage{bbm}
\usepackage{hyperref}
\usepackage{multicol}
\usepackage{listings}
\usepackage{float} 
\usepackage{tabularx}
\usepackage{amsmath}
\usepackage{multirow}
\usepackage{graphicx}
\usepackage{booktabs}
\usepackage{caption}
\usepackage[utf8]{inputenc}

\usepackage{listings}
\usepackage{xcolor}

\definecolor{codegreen}{rgb}{0,0.6,0}
\definecolor{codegray}{rgb}{0.5,0.5,0.5}
\definecolor{codepurple}{rgb}{0.58,0,0.82}
\definecolor{backcolour}{rgb}{0.95,0.95,0.92}
\lstdefinestyle{Tydi}{
    language        =   C++,
    basicstyle      =   \ttfamily\footnotesize,
    keywordstyle    =   \color{magenta},
    stringstyle     =   \color{codepurple},
    commentstyle    =   \color{codegreen},
    breaklines      =   true,
    columns         =   fixed,
    basewidth       =   0.5em,
}

\begin{document}

\frontmatter

\title[tudelft-white]{Tydi-lang: a language for \\ typed streaming hardware}
\subtitle[tudelft-black]{A manual for future Tydi-lang compiler developers}
\author[tudelft-white]{Yongding Tian}
\affiliation{Technische Universiteit Delft}
\coverimage{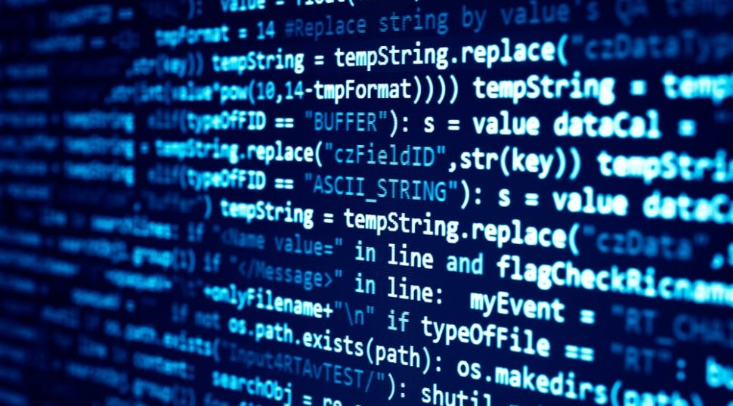}
\covertext[tudelft-white]{
    \vfill
}
\setpagecolor{tudelft-cyan}
\makecover[split]

\input{title}

\chapter*{Abstract}
\setheader{Abstract}
Transferring composite data structures with variable-length fields often requires designing non-trivial protocols that are not compatible between hardware designs. When each project designs its own data format and protocols the ability to collaborate between hardware developers is diminished, which is an issue especially in the open-source community. Because the high-level meaning of a protocol is often lost in translation to low-level languages when a custom protocol needs to be designed, extra documentation is required, the interpretation of which introduces new opportunities for errors. 

The Tydi specification (Tydi-spec) was proposed to address the above issues by codifying the composite and variable-length data structures in a type and providing a standard protocol to transfer typed data among hardware components. The Tydi intermediate representation (Tydi-IR) extends the Tydi-spec by defining typed interfaces, typed components, and connections among typed components.

In this thesis, we propose Tydi-lang, a high-level hardware description language (HDL) for streaming designs. The language incorporates Tydi-spec to describe typed streams and provides templates to describe abstract reusable components. We also implement an open-source compiler from Tydi-lang to Tydi-IR. We leverage a Tydi-IR to VHDL compiler, and also present a simulator blueprint to identify streaming bottlenecks. We show several Tydi-lang examples to translate high-level SQL to VHDL to demonstrate that Tydi-lang can efficiently raise the level of abstraction and reduce design effort.

\input{preface}

\tableofcontents

\mainmatter

\include{01-introduction}
\include{02-background}
\include{03-tydi-language}
\include{04-tydi-language-front-end}
\include{05-tydi-simulator}
\include{06-evaluation-result}
\include{07-conclusion}

\appendix
\include{08-appendix}

\bibliography{ref}

\end{document}

%% file: title.tex
\begin{titlepage}

\begin{center}


{\makeatletter
\largetitlestyle\fontsize{64}{94}\selectfont\@title
\makeatother}

{\makeatletter
\ifx\@subtitle\undefined\else
    \bigskip
   {\tudsffamily\fontsize{22}{32}\selectfont\@subtitle}    
\fi
\makeatother}

\bigskip
\bigskip

by

\bigskip
\bigskip

{\makeatletter
\largetitlestyle\fontsize{26}{26}\selectfont\@author
\makeatother}

\bigskip
\bigskip

to obtain the degree of Master of Science

at the Delft University of Technology,

to be defended publicly on Tuesday July 5, 2022 at 9:00 AM.

\vfill

\begin{tabular}{lll}
    Student number: & 5355710 \\
    Project duration: & \multicolumn{2}{l}{November 1, 2021 -- July 5, 2022} \\
    Thesis committee: & \mbox{Dr.\ Zaid\ Al-Ars}, & Technische Universiteit Delft, supervisor \\
    & \mbox{Prof.\ Peter\ Hofstee},  & Technische Universiteit Delft \\
    & \mbox{Dr.\ Nick\ van\ der\ Meijs}, & Technische Universiteit Delft \\
    & \mbox{Dr.\ Johan\ Peltenburg},  & Technische Universiteit Delft \\
\end{tabular}

\bigskip
\bigskip

\bigskip
\bigskip
An electronic version of this thesis is available at \url{http://repository.tudelft.nl/}.


\end{center}

\begin{tikzpicture}[remember picture, overlay]
    \node at (current page.south)[anchor=south,inner sep=0pt]{
        \includegraphics{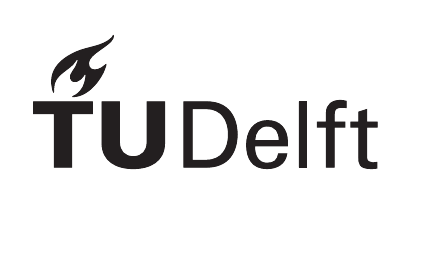}
    };
\end{tikzpicture}

\end{titlepage}

%% file: preface.tex
\chapter*{Preface}
\setheader{Preface}


First of all, I would like to express my gratitude to my supervisors, Peter Hofstee and Zaid Al-Ars, for their suggestions and deep insights in hardware design. Furthermore, I would like to thank Matthijs A. Reukers for implementing the Tydi-IR backend, and many people who previously worked on the Tydi project, such as Johan Peltenburg, Jeroen van Straten, and Matthijs Brobbel. Their contribution to the Tydi specification is the foundation of Tydi-lang. \newline

Secondly, I want to express my Memory of Eelco Visser (1966-2022). He was the head of the Programming Languages Group in TUDelft and the professor who led me to the programming languages field. I recommend all readers to take the Compiler Construction course in TUDelft (if possible) and spend some time on the concepts and ideas in Spoofax \cite{spoofax}. \newline

In addition, I would like to express my deepest appreciation to my parents, Guiyu Tian and Xiaolu Cheng, for their emotional and financial support, as well as their many suggestions for my academic career. \newline

Finally, I would like to thank you, the future Tydi toolchain developer. Tydi-lang, as well as Tydi-IR and Tydi-spec, is a small milestone of a large toolchain. The final toolchain should provide an integrated solution for designing hardware accelerators, from reading host memory to performing calculations on them, with only a few low-level HDL lines (zero is the best). Tydi-lang might also be a generic high-level language for general-purpose streaming hardware design in the future, but that is a longer road. \newline

During my bachelor's study, a professor told me that we write code today because we will not write it again tomorrow. This sentence also describes the Tydi project: we design Tydi today because we do not need to struggle with designing complicated hardware tomorrow. \newline

\begin{flushright}
{\makeatletter\itshape
    \@author \\
    Delft, June 2022
\makeatother}
\end{flushright}

%% file: 01-introduction.tex

\chapter{Introduction}




\section{Context}
In the last ten years, the rate of performance improvements in general purpose processors has not been able to keep up with the growth rate of data. Researchers and engineers proposed several solutions. The first solution uses heterogeneous computing devices such as graphic process units (GPUs) and field-programmable gate arrays (FPGAs). A GPU can accelerate highly parallel operations such as matrix operations due to its execution unit (SM unit in CUDA) uses single-instruction-multiple-data (SIMD) design. 
Nvidia developed the CUDA toolkit \cite{cuda} to provide a general and abstract interface to design GPU-accelerated applications. An FPGA can directly use gate arrays to compute the result without using traditional load-store architecture. FPGAs can be effective accelerators in memory-intensive applications \cite{tens_of_gigabytes, parquet_fpgas, genomics_fpgas} because CPUs need several stages, such as fetching data, fetching instructions, and execution, to perform computations, while FPGAs can directly use logic gates to compute and store the values~\cite{bigdata_fpgas}. 
However, the clock frequency of FPGAs is much lower than the frequency of most CPUs. The frequency advantage on CPUs can compensate for the disadvantages in CPUs' fetch-execution structure. GPUs usually have lower frequencies than CPUs in order to control the power introduced by parallelizing the execution unit. The memory for GPUs is also designed with higher bandwidth but higher latency compared with memory for CPUs. Above all, the acceleration performance of hardware is very application-specific. Hardware designers usually need to consider algorithm parallelism, arithmetic intensity, and frequency.  
Meanwhile, designing an FPGA-based accelerator consumes much more time than designing GPU-based acceleration algorithms because the hardware design flow is much longer than the software design flow. There are also more abstraction layers in the software area, such as abstraction layers for processors with different instruction sets and abstraction layers for different operating systems. However, the abstraction level for designing hardware is much lower. For example, many hardware design tools are vendor-specific, and many IP cores are platform-specific. The second method of catching up with the growth rate of data is using computer clusters to process large-scale datasets. For example, in big-data analytics, Apache Spark \cite{spark} can automatically distribute the data and computing algorithms to multiple compute nodes and manage the data shuffling among cluster nodes. Later, developers found that shuffling data in the network usually causes too much performance overhead because serializing the memory data to data streams and de-serializing back to memory data cost too much time. To address this issue, Apache Arrow \cite{arrow} has been proposed to provide a column-based memory format 
with zero serialization cost. Considering the missing support for FPGA in Arrow, Fletcher \cite{fletcher, fletcher_internals} has been proposed as a tool to automatically generate the hardware interface to access the Arrow memory data for FPGA accelerators. However, the authors of Fletcher realized that representing the nested data structure with hardware description language is inefficient and greatly increases the design complexity. Thus, they proposed Tydi-spec \cite{tydi-spec} which provides a standard for representing hierarchical data 
and the corresponding hardware-level streaming protocols. The hardware designed with Tydi-spec is also called typed streaming hardware because of the built-in type system. 

This thesis \footnote{Some contents of this thesis also appear in ``Tydi-lang: A User-friendly Language for Typed Streaming Hardware'', a paper which I submitted to ICCAD 2022.} reports on designing a programming language based on the Tydi-spec and its corresponding compiler implementation. The Tydi project supervisors extended the Tydi-lang context from the FPGA acceleration area to the general hardware design area, where transferring hierarchical and variable-length data among hardware components often requires designers to manually design protocols and document the protocol specification. This common approach increases the design effort because designers need to frequently switch between reading documentation and writing low-level HDL code, which also introduces new opportunities to make mistakes. In the open-source community, this issue becomes more serious because each project usually designs their own protocols and representations. The standard data representation and protocols to transfer the data in Tydi-spec can efficiently address the issues mentioned above. Because general hardware designing is a much wider area than FPGA accelerators, I will use general hardware as the use case for Tydi-lang in the elaboration and use the FPGA accelerator as the use case in the results and evaluation section. 

\section{Challenge}

The primary aim of this MSc project is to design a Tydi-spec based hardware description language (HDL) for hardware developers. However, designing a language is not as simple as directly presenting the Tydi-spec elements. Designing a language should include considerations such as reducing the hardware design effort, 
and raising the level of abstraction. In addition, hardware verification is also an important phase in the hardware design flow. Performing high-level hardware verification to find out high-level design errors (such as type mismatch) and making the new language compatible with existing tools and simulators are also important. 

Implementing the compiler is another challenge because compilers are usually complicated. The code written by developers is completely unpredictable, but the compiler must be abstract enough to process these unpredictable inputs. One of the authors in Tydi-spec also highly recommended using the Rust language to implement the prototype. The "immutable/mutable reference" feature in Rust also requires new patterns to construct the compiler structure.

\section{Problem statement and research questions}
\label{subsection:problem_statement}

This thesis aims to design a high-level hardware description language for typed streaming hardware and implement the corresponding compiler prototype in Rust. Thus, the research questions can be formulated as below:

\begin{itemize}
    \item What is the essential language syntax to describe typed streaming hardware based on the Tydi-spec?
    \item How to minimize the design effort for Tydi-lang users?
    \item Rust is a relatively new language, its unique immutable/mutable reference system requires more design effort on the memory structure. How can we address the memory challenges specific to designing a compiler in Rust? 
    \item What kind of abstraction method should the compiler provide to facilitate designing typed streaming hardware?
    \item Hardware verification is an important phase in design flow, how to assess hardware verification in the context of a Tydi-spec based toolchain.
    \item How to enable the cooperation between the new language and other existing HDLs and tools?
\end{itemize}


\section{Contributions}
The major contributions of this thesis can be summarized into following points:
\begin{itemize}
    \item Design a user-friendly, type-safe and high-level HDL for streaming hardware and implement its compiler (Tydi-lang).
    \item Introduce the "template" concept for typed streaming hardware.
    \item Provide a new toolchain (Tydi tools and Fletcher) to design FPGA accelerators for big data applications efficiently. This use case might be a foundation for a future trans-compiler from software programming languages to hardware description languages.
    \item Present a high-level simulator blueprint to facilitate design analysis, including identifying streaming bottlenecks, and generate testbenches for low-level verification tools. 
\end{itemize}

\section{Outline}
The remainder of this thesis is organized as follows:

\begin{itemize}
    \item \textbf{Chapter 2: Background} provides relevant background information for the projects discussed in subsequent Chapters.
    \item \textbf{Chapter 3: Tydi language} introduces the syntax and concepts of the Tydi language.
    \item \textbf{Chapter 4: Tydi language compiler frontend} explains how the Tydi compiler is constructed and describes its inner features such as multi-file analysis and multi-threaded compiling.
    \item \textbf{Chapter 5: Tydi simulator} proposes a blueprint for performing packet-level simulation and testbench generation for Tydi-lang.
    \item \textbf{Chapter 6: Result and evaluation} shows some Tydi sample applications which translate selected SQL benchmark queries in TPC-H to the hardware streaming logic. These sample applications demonstrates that the Tydi language can greatly reduce the complexity of developing big data acceleration on FPGAs.
    \item \textbf{Chapter 7: Conclusion} summarizes the thesis.
    \item \textbf{Chapter 8: Appendix} records some extra but non-trivial information for the Tydi compiler project.
\end{itemize}

%% file: 02-background.tex
\chapter{Background}




\section{Tydi Specification (Tydi-spec) and Tydi intermediate representation (Tydi-IR)}
As mentioned earlier, Tydi-spec provides a standard method for describing hierarchical data structures using combinations of logical types and defines how to map the data to hardware streams. There are a total of five logical types: \texttt{Null}, \texttt{Bit}, \texttt{Group}, \texttt{Union} and \texttt{Stream} in Tydi-spec. To describe the type system in Tydi-spec, Tydi-IR is proposed as an intermediate representation to encode the logical types directly. In addition, Tydi-IR also extends Tydi-spec with some hardware-level concepts such as \texttt{Port}, \texttt{Streamlet}, \texttt{Implementation}, \texttt{Connection}, and \texttt{Instance}. These concepts can efficiently describe typed components and circuits. Table \ref{table:tydi_ir_terms} summarizes the terms in Tydi-spec and Tydi-IR.

\begin{table}[]
\centering
\caption{Terms used in Tydi-spec and Tydi-IR}
\label{table:tydi_ir_terms}
\resizebox{\textwidth}{!}{
\begin{tabular}{|c|c|l|}
\hline
Term & Type & \multicolumn{1}{c|}{Meaning} \\ \hline
Null & Logical type & Represents empty data. A stream of null type will be optimized out. \\ \hline
Bit(x) & Logical type & Represents data that requires x hardware bit to represent. \\ \hline
Group(x,y) & Logical type & \begin{tabular}[c]{@{}l@{}}A tuple of several other logical types (x and y in this example). The total hardware bit \\ would be the sum of all child element bit width.\end{tabular} \\ \hline
Union(x,y) & Logical type & \begin{tabular}[c]{@{}l@{}}An union of several other logical types (x and y in this example). The total hardware bit\\ would be the maximum bit width of a single child.\end{tabular} \\ \hline
Stream(x) & Logical type & \begin{tabular}[c]{@{}l@{}}Represents a stream of a logical type. The stream can also specify the data dimension,\\ protocol complexity, hardware synchronicity, and throughput as optional arguments.\end{tabular} \\ \hline
Port & Hardware element & Represents a hardware port, the port must specifies its logical stream type and direction. \\ \hline
Streamlet & Hardware element & \begin{tabular}[c]{@{}l@{}}Represents the port map of a component. This term is almost the same as the "entity" \\ term in VHDL.\end{tabular} \\ \hline
Implementation & Hardware element & \begin{tabular}[c]{@{}l@{}}Represents the inner structure of a component. The inner structure should be a combin-\\ ation of instances and connections. Implementation must specify a streamlet as its port \\ map, this relationship is similar to the relationship between "entity" and "architecture" in \\ VHDL. Implementation can be declared as "external" if they cannot be represented by \\ instances and connections. "Implementation" is also called "impl" in Tydi-lang. \end{tabular} \\ \hline
Connection & Hardware element & \begin{tabular}[c]{@{}l@{}}Connect two ports. The two ports must have the same data stream type, compatible \\ protocol complexities, correct directions and same clock domain. Connections must be \\ declared in implementation.\end{tabular} \\ \hline
Instance & Hardware element & \begin{tabular}[c]{@{}l@{}}Represents a nested implementation instance in another implementation. The port of the \\ nested  implementation can be accessed by using the instance.\end{tabular} \\ \hline
Clock domain & Hardware Clock & \begin{tabular}[c]{@{}l@{}}A clock domain is a representation of clock frequency and phase and is usually bound to \\ a port. Due to the handshaking mechanism in the stream, the clock domain concept \\ ensures only two ports with the same clock domains can be connected together. \end{tabular} \\ \hline
\end{tabular}%
}
\end{table}

For an example of the logical type system, suppose we want to represent a RGB pixel whose color depth is 8 bits. We can define three logical types and each of them is represented by a \texttt{Bit(8)}, and define a logical group, called \texttt{Pixel}, to combine the three channel with \texttt{Group(red,green,blue)}. The \texttt{Pixel} will map to 24 hardware bits now but it is not stream yet. We can use \texttt{Stream(Pixel)} to define a stream of \texttt{Pixel} data. \texttt{Stream} is also a logical type and can be put in another stream with \texttt{Stream(Group(Stream))}. In Tydi-spec, this case is called nested stream and some stream properties can describe the stream behaviors and relationships. These properties are listed below:

\begin{itemize}
    \item Dimension: describes the dimension of the data. For example, an English character is a 0-dimension ASCII stream, a word is a 1-dimension stream and a sentence is a 2-dimension stream. 
    \item User: a logical type to deliver bit-oriented data rather than stream-oriented data.
    \item Throughput: represents the minimum number of elements that should be transferrable on the child stream per element in the parent stream.
    \item Synchronicity: represents the relation between the dimensionality information of the parent stream and the child stream.
    \item Complexity: this is a number to represent the complexity for the corresponding physical stream interface.
    \item Direction: represents the direction of the stream. The direction for nested stream is set relative to its parent stream. For example, defining a reverse stream A in a nested Stream B which is also reversed results in a forward stream A.  
    \item Keep: represents whether the stream carries extra information beyond the "stream" and "user" payload and whether to keep this stream when both carried data and user data are \texttt{Null}.
\end{itemize}

Though Tydi-IR provides abilities to describe typed components and circuits, it is still very different from the Tydi-lang. Tydi-IR, like many other intermediate representations, is usually too long and contains excessive extra information, and thus is not suitable for developers. Many high-level features are also not designed in Tydi-IR such as the syntax for variables/for/if.

The Tydi-IR project is done by Matthijs A. Reukers as his MSc thesis. Readers should be able to find his thesis in the TUDelft database to find more details.

\section{C++ compiler and Rust compiler}
The two compilers are mentioned here because some of their features and design decisions are referenced in the Tydi-lang compiler. The two most obvious reference points are the \texttt{typename} keywords in C++ and the multi-file analyzing mechanism in Rust.

\subsection{\texttt{typename} keyword in C++}
In C++ syntax, \texttt{typename} is used in declaring a template and in using a type identifier as template argument as shown in the following code snippets.

\begin{lstlisting}[language=c++]
class text
{
public:
	class print_interface
	{
	public:
		void print()
		{
			std::cout << "this is a text" << std::endl;
		}
	};
	
};

template <typename T>
class invoke_print
{
public:
	static void print()
	{
		using printable = typename T::print_interface;
		printable temp;
		temp.print();
	}
	
};

int main()
{
	invoke_print<text>::print();
	return 0;
}
\end{lstlisting}

The \texttt{typename} At line 15 indicates the class has one arguments which must be a type. The \texttt{typename} at line 21 indicates the term \texttt{T::print\_interface} should be evaluated as a type. C++ developers can define a global variable and a class with the same name in the single source file because variables and types are stored in two separate space. This feature also makes binding a type to a variable ambiguous because the compiler cannot determine whether the identifier refers to a type or a variable. Thus C++ uses the \texttt{typename} keyword to clarify it.

The Tydi-lang also has similar mechanisms to clarify the meaning of the identifier in templates. Because there are streamlets and implementations, the keywords applied in Tydi-lang templates are more complicated than those in C++. I tried to provide something easier than the current keywords system, for example, making "type", "streamlet", "impl" keywords optional (because it is counter-intuitive for users to write "type" keywords before types), but unfortunately failed due to some parser limitations that are discussed in Section \ref{subsubsection:limitations_for_pest}.  

\subsection{Multi-file analysis in Rust}
In C++, the source codes are split into source files (.cpp files) and header files (.h files). Header files only declare the identifier, the arguments, and the return type of functions. Source files define the implementation of these functions. During compilation, the content of the header files will be copied to the position of the \texttt{include "header"} statement in source files to ensure the functions invoked in a source file are always defined. After compiling, the C++ linker will link the function implementations defined in multiple source files.

With the significant improvements in C++ standard template library (STL) introduced in C++ 11, many developers have begun to mix the source files and the header files because separating them for templates is complicated. This trend results in a new file type, called C++ header files (.hpp), and the renaming of the previous header files to C-compatible/pure-C header files (.h). However, developers found poor compiling performance when they used many C++ header files in a project with only a few source files. The cause is that the C++ compiler can allocate a maximum of one thread for a single source file. When a single source file becomes longer and longer as more header files are included, the compiler spends more and more time to compile it. Thus a side effect of using the template is to prevent multi-threaded compilation.

Rust, as a newly developed programming language, has an entirely different mechanism to handle function definitions and implementations to achieve multi-threaded compilation. The major difference compared with C++ is that the C++ compiler will check the function definition when a function is invoked. In contrast, the Rust compiler will collect all the definitions first and then resolve them to their definitions later. 
This difference also explains that invoking a function defined after using is not allowed in C++ but acceptable in Rust. The following code snippets show this difference.

\begin{lstlisting}[language=c++]
//C++ version
int main()
{
	print(); //error, print not defined
	return 0;
}

void print()
{
	std::cout << "Hello world" << std::endl;
}
\end{lstlisting}

\begin{lstlisting}[language=c++]
//Rust version
fn main() {
    test(); //no error
}

fn test()
{
    println!("Hello, world! {}", 2);
}
\end{lstlisting}

The Tydi-lang compiler uses a similar mechanism to Rust to support multi-file compiling.

\section{Apache Arrow and Fletcher}
Apache Arrow and Fletcher are mentioned here because they can integrate with Tydi-lang as a large toolchain to design hardware accelerators as shown in Chapter \ref{chapter:result_and_evaluation}.

Apache Arrow \cite{arrow} is a widely-applied data format with zero serialization overhead in the big data analytic area. The in-memory representation for Apache Arrow data is column-based (the data addresses in the same column are continuous in memory). This feature accelerates Map-Reduce operations by enabling prefetching data on the processor level. A limitation of this feature is that Apache Arrow has chosen to make these tables immutable because of the large overhead of inserting new data into a continuous memory region. 

Fletcher \cite{fletcher} is an open-source framework to automatically generate hardware components for FPGA accelerators to access Arrow data in host memory. The generation of components is based on the Arrow data schema, and the hardware connection between the FPGA and the host processor could be PCI-E or OpenCAPI. Hardware designers can design their acceleration algorithms by other tools, such as High-Level Synthesis or OpenCL, and let the acceleration algorithms work with Fletcher to process the memory data with FPGA accelerators. 

%% file: 03-tydi-language.tex
\chapter{Tydi language}

\section{Introduction to Tydi language (Tydi-lang)}
\label{subsecion:The_aim_of_Tydi_language}
Tydi-lang is a high-level hardware description language based on the type system introduced in the Tydi-spec \cite{tydi-spec} and Tydi-IR. This new language aims to raise the level of abstraction for typed streaming hardware and reduce the design effort for hardware designers. 

\begin{figure}
    \centering
     \includegraphics[width=\textwidth]{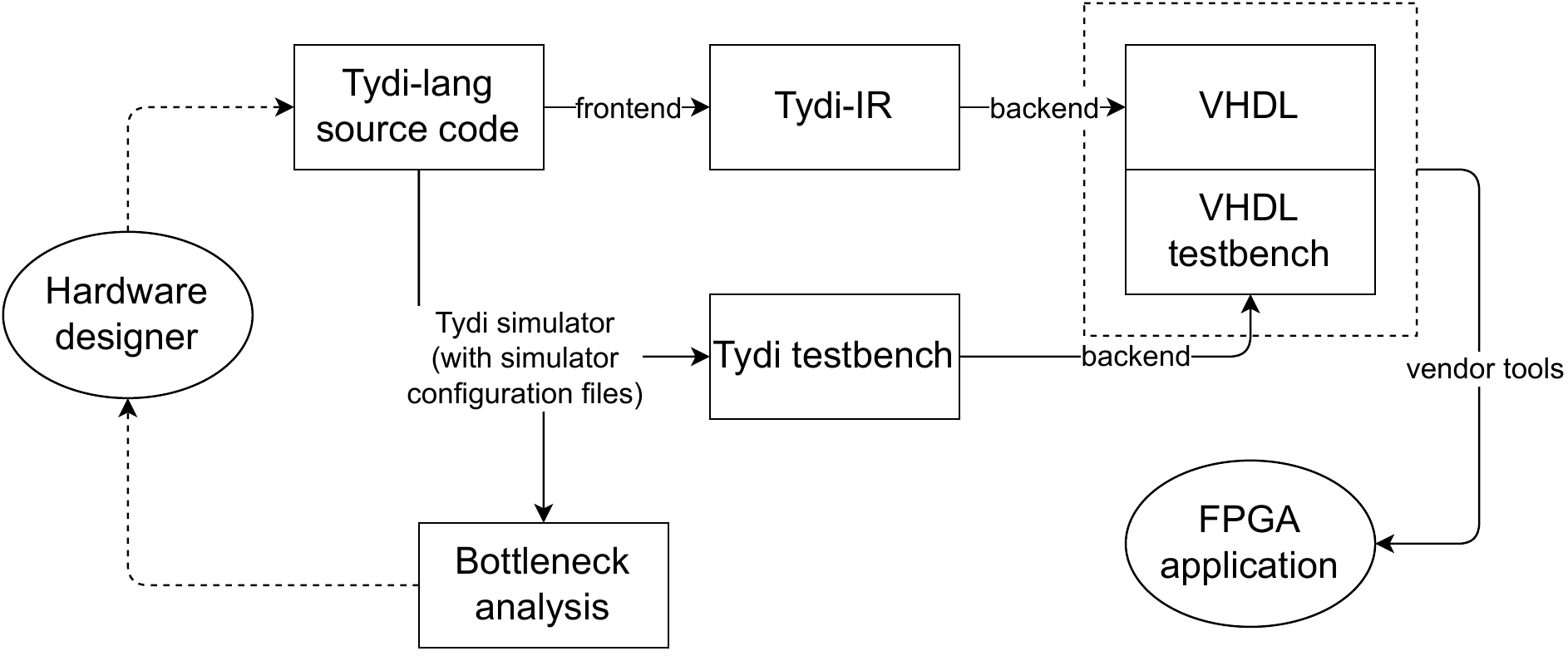}
    \caption{Tydi-lang toolchain workflow}
    \label{fig:tydi_lang_tool_chain_work_flow}
\end{figure}

Figure \ref{fig:tydi_lang_tool_chain_work_flow} provides an overview for Tydi-lang toolchain. The Tydi-lang source code can be compiled to Tydi-IR with the Tydi-lang compiler and further compiled to VHDL with a compiler introduced in the Tydi-IR paper. The Tydi-lang compiler is also called "a frontend of Tydi" because the output is an intermediate representation. Similarly, the compiler in the Tydi-IR paper is called "a backend of Tydi" because it compiles Tydi-IR to VHDL. In the future, we plan to introduce other frontends and backends to allow interactions with other toolchains such as CHISEL \cite{chisel}. The simulator configuration files contains the input data sequence, clock information and the name of the top-level implementation.

\begin{figure}
    \centering
     \includegraphics[width=\columnwidth]{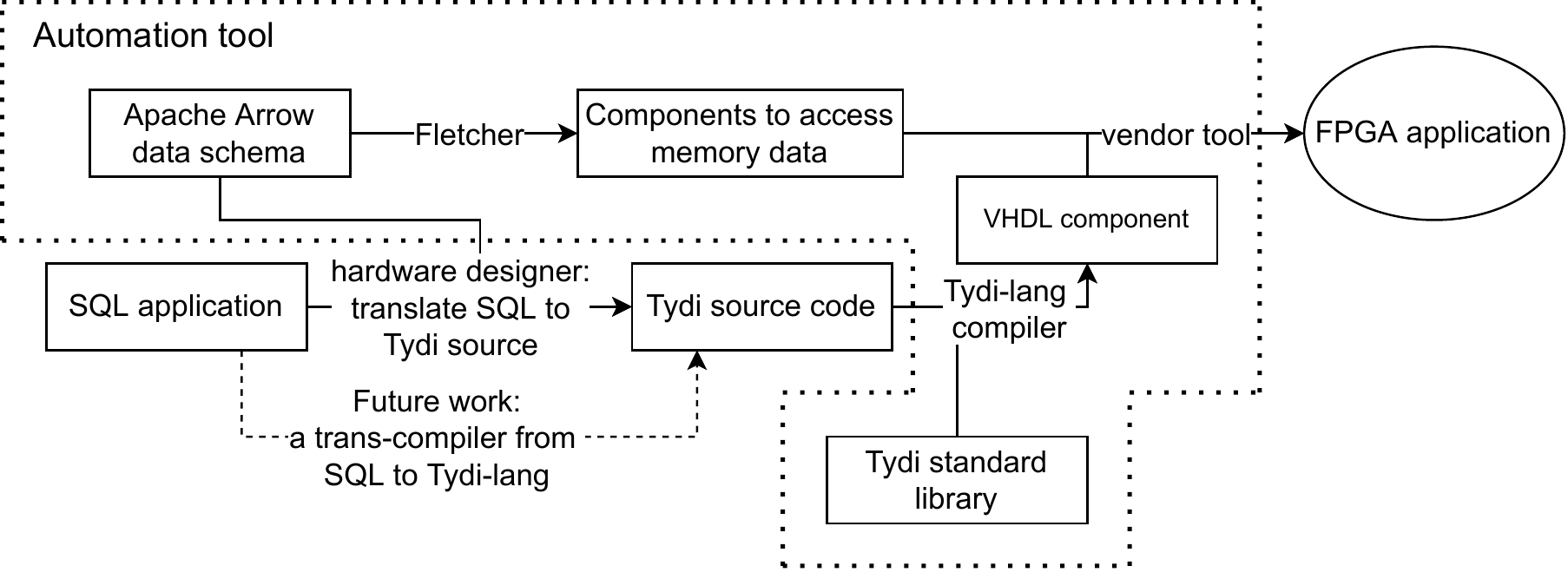}
    \caption{Tydi-lang workflow in big data}
    \label{fig:tydi_lang_in_big_data}
\end{figure}

To illustrate the use of Tydi we provide an example from big data, Tydi-lang can be an elegant bridge to connect query languages and the FPGA accelerators, as shown in Figure \ref{fig:tydi_lang_in_big_data}. Big data developers usually use SQL to do analytics on a dataset with a known schema. We use Apache Arrow as the dataset format because it is widely applied in big data applications for zero serialization overhead. With Fletcher \cite{fletcher}, which is a tool to generate hardware components to access Apache Arrow data automatically, the design effort can be greatly reduced while the only thing left to do is translating the SQL to Tydi-lang. Our experience suggests it is possible to design a tool to automatically compile SQL to Tydi-lang in the future.

Based on Tydi-spec and Tydi-IR, Tydi-lang introduces a generative syntax and a template concept, which allows developers to describe hardware components in a more abstract and reusable way. These two features also allow developers to design streaming hardware more efficiently by directly connecting components at a higher level and facilitate translating software languages to Tydi-lang. Some frequently-used component templates are introduced in a standard library for Tydi-lang. One of the benefits of using the Tydi-lang standard library is that developers can design digital circuits without having to use low-level HDLs, for example to accelerate SQL queries via FPGA accelerators, where operations on data can be mapped to hardware templates. Besides the standard library, the Tydi-lang also integrates a high-level design rule check system to identify type errors, which would be un-trackable on the lower layer.

Based on the Tydi-lang, a simulator is proposed to assist high-level developers in designing their streaming circuits to meet functional requirements regardless of low-level behavior. In traditional hardware design, developers need to care about optimizing low-level components and meeting high-level functional requirements at the same time. The change in the delay time of low-level components may cause different high-level throughput because the arrival time of asynchronous input data packets determines the delay. Analyzing the timing information of all components can quickly help developers find the streaming bottlenecks. Traditional low-level simulators can also be used to find bottlenecks but doing so is cumbersome. For some cases where the low-level components have not been designed, high-level developers can use theoretical data in simulation and continue working without waiting for the completion of the low-level side.

\section{Tydi language features}
This section describes the Tydi-lang features from a theoretical aspect, which has already been discussed in the Tydi-lang paper - "Tydi-lang: A User-friendly Language for Typed Streaming Hardware".

\subsection{High level design}
\label{subsubsection:High_level_design}
Tydi-lang is designed to be a high-level hardware description language. The term "high-level" does not refer to high-level synthesis or something similar. However, it refers to the notion that the developers do not have to care about low-level properties of the component, such as delay, circuit area, and clock frequency. Developers only focus on meeting the functional requirements of the circuit. In the big data analytics area, the functional requirements of big data applications constantly change due to many aspects, such as market changing, new types of collected data, new decisions from companies. To meet these high-level requirements, we need Tydi-lang to design prototype circuits effectively and avoid struggling with low-level components.

The high-level feature of Tydi-lang also makes it possible to directly map software data structures and code patterns to logical types and patterns in Tydi-lang. 
For example, software programming languages have two types of "for" loops. The two types are "for" loops without shared values and with shared values. The "for" loops without shared values can be accelerated by parallelizing the computation of a single loop. This pattern can be easily achieved with the generative "for" syntax in Tydi-lang, which generates parallel hardware components and connections.

However, due to the high-level feature, Tydi-lang is not efficient in describing the behaviors of low-level components. Components that can be described as connections and instances are not considered low-level components. For example, describing the behavior of an adder is hard in Tydi-lang because the adder itself is already extremely basic. In practical Tydi-lang projects, these low-level components' behavior can be written using traditional HDLs or CHISEL, and the structure code and behavior code can be merged in the synthesis stage.  

\subsection{Hardware description by variables}
\label{subsubsection:Hardware_decription_by_variables}
Please notice that the variable here refers to the variable system in Tydi-lang rather than the "variable" in VHDL. In traditional VHDL, developers need to manually specify the properties of each port, such as port width and the number of ports. In Tydi-lang, the port width is represented by logical types, and an integer can characterize the number of ports. This feature is powerful for mapping readings data multiple times in software code to Tydi-lang. Let us suppose we have a piece of SQL, and it uses a variable twice. In software, there is nothing wrong because each variable is a value in a register or memory and can be accessed twice. However, the data is transient in hardware because it is stored in logical gates. In streaming hardware, there should be a component to duplicate the data packet and send them to two ports. The duplicator will not acknowledge the source component until both sink components acknowledge that the data packets are received. However, if the variable is used three times, there should be three output ports. There should be different numbers of output ports under different cases, and in Tydi-lang, the solution is to use variables to describe hardware. There are totally five types of basic variables (introduced in Section \ref{subsubsection:constant_variable}) which should be enough to describe a hardware component.

Another useful case for applying variables is that developers can easily customize components. For example, we have a constant data generator that always sends a packet whose value is 50 and whose logical type is \texttt{Bit(8)}. The value and the logical type can be declared as two arguments of the component.

There are also many basic cases of applying variables to describe hardware. For example, calculating the minimum bit length to represent a value in memory. The following expression can be used to represent a value in range [$0$, $10^{15}$]: \texttt{Bit(ceil(log2(10$\wedge$15-1)))}.

\subsection{Abstract hardware templates}
\label{subsubsection:Abstract_hardware_templates}
The template concept is one of the most important features in Tydi-lang. Tydi-lang is a strict-type language, and developers might need to define multiple components for different logical types even though the behavior of these components is completely identical. For example, in SQL, \texttt{Decimal(32bit,2)} and \texttt{Decimal(32bit, 4)} are two different logical types because the digit sizes after the decimal point are different. Their corresponding adders are the same on the low-level side because both types are 32 bit for the adder. In this case, the adder should be described as an abstract component because the logical type does not determine its functional correctness.

Abstract hardware templates can also be used to describe some components that are independent of logical types. For example, there is a special component called "voider" in Tydi-lang, which is used to acknowledge the hand-shaking signals of unused ports to avoid blocking other data transmissions. The voider only works on the hand-shaking wires, but the Tydi type system forces it to have a logical type even though the voider will never work on these data wires. Here we can declare the voider as a template, and it can work on many different logical types.  

\subsection{Generative syntax}
\label{subsubsection:Generative_syntax}
Generative syntax means some components are automatically generated from Tydi-lang source code rather than described by developers. The following language syntax and procedures will generate new components for the circuit.

\begin{itemize}
    \item "For" syntax will automatically generate parallel components.
    \item "If" syntax will generate certain components according to a boolean variable.
    \item The instantization of templates will generate new components.
    \item The sugaring process (mentioned in Section \ref{subsubsection:sugaring}) 
    will automatically generate new components for unconnected ports and ports that are used for multiple times.
\end{itemize}

\section{Tydi language specification}
\label{section:tydi_language_specifiction}
This section will explain the Tydi language syntax and concepts, and how these concepts are related to the aims mentioned in Section \ref{subsecion:The_aim_of_Tydi_language}. The relationships between aims and concepts are listed below for fast content locating purpose.

\begin{itemize}
    \item \textbf{Section \ref{subsubsection:High_level_design} High level design} and \textbf{Section \ref{subsubsection:Hardware_decription_by_variables} Hardware description by variables} are related to almost all contents in this section.
    \item \textbf{Section \ref{subsubsection:Abstract_hardware_templates} Abstract representation of hardware} is related to \textbf{Section \ref{subsubsection:Template} Template}.
    \item \textbf{Section \ref{subsubsection:Generative_syntax} Generative syntax} is related to \textbf{Section \ref{subsubsection:If_and_for_block} If and for block} and \textbf{Section \ref{subsubsection:Template} Template}.
\end{itemize}

\subsection{Comments and space}
There are two types of comments in Tydi-lang: line comments and block comments. Their syntax is identical to the corresponding syntax in C++. Tydi-lang is case-sensitive, keywords must have the correct capitalization, and identifiers with capital and non-capital letters are different. Tydi-lang is space-insensitive, and "Space" or "Tab" characters make no difference to the compiler output. Though there is no restriction on the code format, the format used in this thesis is recommended. The following code snippet gives some comment examples

\begin{lstlisting}[language=c++]
//this is a line comment
/*this is a block comment*/
/* 
this is 
also a 
block comment
*/
\end{lstlisting}

\subsection{Scope and name resolution}
Briefly speaking, the scope and name resolution system in Tydi-lang is similar to the one in C++. Both languages use brackets to define a new scope, and language elements in the inner scope can access elements in the outer scopes. Here, the language elements can be variables, classes in C++ and variables, logical types, streamlets, and implementations in Tydi-lang. Accessing target members such as ports and implementation instances in Tydi-lang is similar to accessing class members in C++. However, there are also many differences between the two languages. The detailed scope and name resolution rules (and the difference from C++) are presented in the following paragraphs.

The term "name" means an identifier in code. Name resolution means the process of finding the definition of names. In Tydi-lang, a name can be a combination of English characters, underscore, and numbers but must start with an English character or underscore. Consecutive underscores are not allowed because the Tydi-lang compiler back-end uses consecutive underscores as hierarchy separators.

In Tydi-lang, "scope" is a space to define all language elements such as constant variables, logical types, etc. A scope can be defined in another scope where the two scopes are connected by a directed linkage called "scope relationship". For each Tydi-lang source file, a file itself is a top-level scope that contains many other inner scopes. A Tydi-lang project can contain multiple Tydi-lang source files, and each file must define its package name in the first line. Here is a Tydi-lang language example.

\begin{lstlisting}[language=c++]
package tpch;
type bit5 = Bit(5);
type Group Date {
  year: Bit(32),
  month: Bit(4),
  day: bit5,  //access a logical type in external scope
};
\end{lstlisting}

\begin{figure}
    \centering
    \includegraphics[width=0.5\columnwidth]{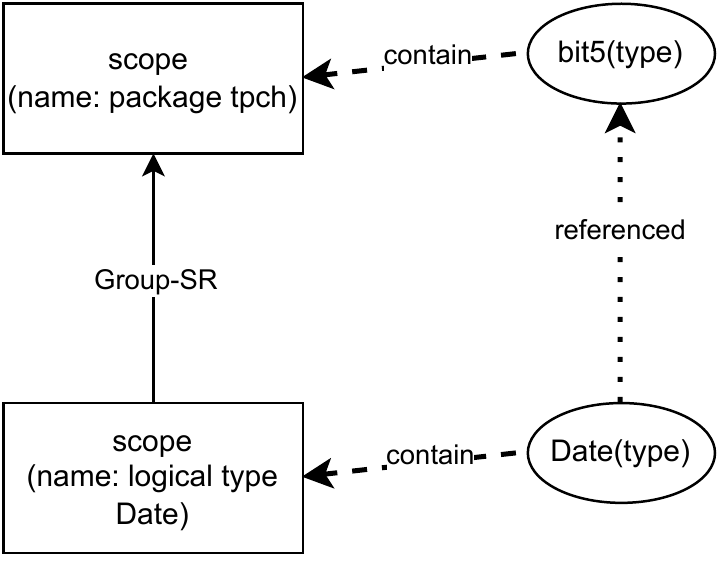}
    \caption{The scope graph representation}
    \label{fig:scope_graph_0}
\end{figure}

The corresponding scope graph is provided in Figure \ref{fig:scope_graph_0}. The above code creates a logical group type called "Date". The logical type "Date" crates a new scope that contains three inner types. These three inner logical types should be able to access external logical types but should be inaccessible from the external scope. So in Tydi-lang, the inner scope "Date" will have a directed relationship to the external file scope. When the Tydi-lang compiler performs name resolution, it can automatically go through scope relationships if the name is not found in the current scope. This mechanism also introduces "name shadowing" that an inner name can shadow an external name.

In most software programming languages, users cannot access a variable before it is declared due to the register allocation. However, in Tydi-lang, names can use other elements without declaring them first once the elements are accessible from the current scope. For example, the following syntax is allowed.

\begin{lstlisting}[language=c++]
package test;
type Group Date {
  year: Bit(32),
  month: Bit(4),
  day: bit5,  //access an external logical type which is not yet defined
};
type bit5 = Bit(5);
\end{lstlisting}

In Tydi-lang, there are 6 types of scope relationships.
\begin{itemize}
    \item \textbf{Group-SR} means the inner scope is created by a logical group type.
    \item \textbf{Union-SR} means the inner scope is created by a logical union type.
    \item \textbf{Stream-SR} means the inner scope is created by a logical stream type. However, this relationship is not frequently used because the properties of a stream is embedded in the stream.
    \item \textbf{Streamlet-SR} means the inner scope is created by a streamlet (explained in Section \ref{subsubsection:Streamlet}).
    \item \textbf{Implement-SR} means the inner scope is created by a implementation (explained in Section \ref{subsubsection:Implement}).
    \item \textbf{IfFor-SR} means the inner scope is created by a "for" statement or "if" statement, this scope relationship is special because it bans most name resolutions across it (explained in Section \ref{subsubsection:If_and_for_block}).
\end{itemize}

The accessibility of language elements after passing scope relationships is described by Table \ref{table:name_resolution_rule}. Please notice that for different language elements, the rules are different. The \texttt{IfFor-SR} will be handled by a generative process rather than an ordinary name resolution process, so it is "not applicable" in the table. In the current Tydi-lang version, connections cannot be referenced by any other language elements, so it is also "not applicable".

Constant variables and logical types are basic building blocks to describe hardware streamlets and implementations, so they are always accessible no matter the scope relationships. In Tydi-lang, streamlets and implementations can only be defined in the package scope because nested streamlets or implementations do not make sense for hardware. Streamlets can pass \texttt{Implement-SR} because declaring implementations need to specify streamlet first. implementations are allowed to pass \texttt{Implement-SR} because declaring instance needs to access other implementations in the external scope. Accessing ports and instances in Tydi-lang is similar to accessing a member in C++. This accessing process must resolve the instance name first and cannot be described by scope relationships, so they are always banned from passing any kinds of scope relationships.

\begin{table}[]
\centering
\caption{Name resolution rule}
\label{table:name_resolution_rule}
\resizebox{\textwidth}{!}{
\begin{tabular}{l|c|c|c|c|c|c}
\hline
 & \multicolumn{1}{l|}{Group-SR} & \multicolumn{1}{l|}{Union-SR} & \multicolumn{1}{l|}{Stream-SR} & \multicolumn{1}{l|}{Streamlet-SR} & \multicolumn{1}{l|}{Implement-SR} & \multicolumn{1}{l}{IfFor-SR} \\ \hline
Const variable & allowed & allowed & allowed & allowed & allowed & N/A \\ \hline
Logical type & allowed & allowed & allowed & allowed & allowed & N/A \\ \hline
Streamlet & banned & banned & banned & banned & allowed & N/A \\ \hline
Port & banned & banned & banned & banned & banned & N/A \\ \hline
Implementation & banned & banned & banned & banned & allowed & N/A \\ \hline
Instance & banned & banned & banned & banned & banned & N/A \\ \hline
Connection & N/A & N/A & N/A & N/A & N/A & N/A \\ \hline
\end{tabular}
}
\end{table}

The corresponding Rust implementation of the scope system is provided in the following link: 
\href{https://github.com/twoentartian/tydi-lang/blob/main/tydi_lang_raw_ast/src/scope.rs}{tydi-lang/tydi\_lang\_raw\_ast/src/scope.rs}

The name resolution rules are defined in other files. For example, the name resolution rule for constant variables is declared in the constant variable file.

\subsection{Constant variable}
\label{subsubsection:constant_variable}

Traditional HDLs focus on precisely and directly describing the hardware. For example, developers use \\ \texttt{STD\_LOGIC} and \texttt{STD\_LOGIC\_VECTOR} to describe the hardware signal directly. However, this hardware representation removes the original human-readable information. For example, in the case of converting Decimal(10,2) in SQL to STD\_LOGIC\_VECTOR(0 to 33) in VHDL, the information "last two digits are after the decimal point" is removed, and the number 33 cannot indicate the range of data, either.

The constant variable system in Tydi-lang is designed to provide a readable, configurable, and abstract way to describe logical types, streamlets, and implementations. There are totally 5 types of constant variable and are presented in Table \ref{table:const_value_type}.

\begin{table}[]
\centering
\caption{Constant variable types}
\label{table:const_value_type}
\resizebox{\textwidth}{!}{
\begin{tabular}{|r|l|}
\hline
Type & \multicolumn{1}{l|}{Meaning} \\ \hline
Integer (int) & represents a 64-bit integer value, example: 1,0b0001,0x01,0o01(octal) \\ \hline
String (str) & represents a string, non-fixed length, example: "forward" \\ \hline
Float (float) & represents a floating number, example: 1.01 \\ \hline
Boolean (bool) & represents a logical value, example: true, false \\ \hline
Clockdomain (cd) & represents a clockdomain, composed of a frequency and a phase. \\ \hline
\end{tabular}%
}
\end{table}

Clockdomain is used to verify that two connected ports have the same clock frequency and phase. Otherwise, the streaming protocol described in Tydi specification \cite{tydi-spec} might not work. The following code illustrates how to declare these constant variables.

\begin{lstlisting}[language=c++]
package test;
const flag: bool = true; //explicit type
const flag = true;
const int0: int = 2;
const int0 = 2;
const str0: str = "hello world";
const str0 = "hello world";
const f0: float = 1.0;
const f0 = 1.0;
const cd0: clockdomain;
const cd1: clockdomain;
const cd2: clockdomain = "100MHz-ph1";
const cd3: clockdomain = "100MHz-ph1";
\end{lstlisting}

The type indicators are optional except for clockdomain values because we want to disambiguate clockdomain expressions from string expressions. The expression of a clockdomain can be empty (in this case, its expression is automatically generated by the Tydi compiler) or a string. In the above code example, cd0 and cd1 are two different clockdomain values and have different random expressions generated by the Tydi-lang compiler. However, cd2 and cd3 are the same because they have the same clockdomain expression.

The Tydi-lang compiler integrates a math engine to evaluate the values of constant variables. 

\begin{table}[]
\centering
\caption{Math operation on constant variables}
\label{table:math_operators}
\resizebox{\textwidth}{!}{
\begin{tabular}{|c|c|c|c|c|c|}
\hline
Operator & Meaning & Operand and output type & Operator & Meaning & Operand and output type \\ \hline
- & unary minus & int-\textgreater{}int / float-\textgreater{}float & \textgreater{}= & is larger or equal & int/float*int/float -\textgreater bool \\ \hline
! & unary not & bool-\textgreater{}bool & \textless{}= & is smaller or equal & int/float*int/float -\textgreater bool \\ \hline
\textless{}\textless{} & bit wise left shift & int*int-\textgreater{}int & \textgreater{} & is larger & int/float*int/float -\textgreater bool \\ \hline
\textgreater{}\textgreater{} & bitwise right shift & int*int-\textgreater{}int & \textless{} & is smaller & int/float*int/float -\textgreater bool \\ \hline
\&\& & logical and & bool*bool-\textgreater{}bool & + & add & \begin{tabular}[c]{@{}c@{}}int*int-\textgreater{}int\\ int*float-\textgreater{}float\\ float*int-\textgreater{}float\\ float*float-\textgreater{}float\\ str+int/float/bool-\textgreater{}str\\ int/float/bool+str-\textgreater{}str\end{tabular} \\ \hline
|| & logical or & bool*bool-\textgreater{}bool & - & minus & \begin{tabular}[c]{@{}c@{}}int*int-\textgreater{}int\\ int*float-\textgreater{}float\\ float*int-\textgreater{}float\\ float*float-\textgreater{}float\end{tabular} \\ \hline
== & equal to & \begin{tabular}[c]{@{}c@{}}int*int-\textgreater{}bool/\\ float*float-\textgreater{}bool/\\ str*str-\textgreater{}bool/\\ cd*cd-\textgreater{}bool/\\ bool*bool-\textgreater{}bool\end{tabular} & * & multiply & same as - \\ \hline
!= & not equal to & same as == & / & divide & \begin{tabular}[c]{@{}c@{}}int*int-\textgreater{}int\\ int*float-\textgreater{}float\\ float*int-\textgreater{}float\\ float*float-\textgreater{}float\end{tabular} \\ \hline
\& & bitwise and & int*int-\textgreater{}bool & \% & modulo & int*int-\textgreater{}int \\ \hline
| & bitwise or & int*int-\textgreater{}bool & \textasciicircum{} & power & same as - \\ \hline
$\sim$ & bitwise not & int-\textgreater{}int & round(x) & math rounding & float-\textgreater{}int \\ \hline
log a(b) & log & int/float*int/float -\textgreater float & floor(x) & math flooring & float-\textgreater{}int \\ \hline
 &  &  & ceil(x) & math ceiling & float-\textgreater{}int \\ \hline
\end{tabular}%
}
\end{table}

With the help of constant variable, the previous decimal example can be converted to the following Tydi-lang code. 

\begin{lstlisting}[language=c++]
package test;
const max_decimal_10 = 10^10 - 1;
const bit_width_decimal_10 = ceil(log2(max_decimal_10));
type SQL_decimal_10 = Bit(bit_width_decimal_10);
type Group SQL_decimal_10_2 {
  const frac = 2,  //other code can access the frac within the logical type "SQL_decimal_10_2"
  decimal: SQL_decimal_10,
};
type SQL_decimal_10_2_stream = Stream(SQL_decimal_10_2, d = 1);
\end{lstlisting}

The type SQL\_decimal\_10\_2 in Tydi-lang is much more flexible and human-readable than direct VHDL. The logical group type also includes the fraction information as a constant variable.

The corresponding Rust implementation of the constant variable system is provided below:

\begin{itemize}
    \item Variable system: \href{https://github.com/twoentartian/tydi-lang/blob/main/tydi_lang_raw_ast/src/variable.rs}{tydi-lang/tydi\_lang\_raw\_ast/src/variable.rs}
    \item Type system (includes complex types such as streamlet and implement):
    \href{https://github.com/twoentartian/tydi-lang/tydi_lang_raw_ast/src/data_type.rs}{tydi-lang/tydi\_lang\_raw\_ast/src/data\_type.rs}
    \item The math system and evaluation of variables: \href{https://github.com/twoentartian/tydi-lang/blob/main/tydi_lang_parser/src/evaluation_var.rs}{tydi-lang/tydi\_lang\_parser/src/evaluation\_var.rs}
\end{itemize}

\subsection{Logical type}
\label{subsubsection:Logical_types}

As aforementioned, there are five logical types in Tydi-spec: \texttt{Null}, \texttt{Bit}, \texttt{Group}, \texttt{Union} and \texttt{Stream}. The ways of defining logical types are illustrated in the following code snippet. Please notice that \texttt{type \{id\} = \{logical\_type\}} is making an alias of a logical type. \texttt{Bit} is a basic logical type so in most cases we only make alias of it.

\begin{lstlisting}[language=c++]
package tpch;

type byte = Bit(8); //define an alias of a Bit(8)
type Group rgb {  //define a group type
  const x = 8,    //it is possible to define const variables in logical group scope
  r: Bit(x),
  g: Bit(x),
  b: Bit(x),
};
type Union rgb_null {
  rgb_data: rgb,
  null_data: Null, //Null logical type
};
type rgb_null_alias = rgb_null;

type rgb_stream = Stream(rgb);
\end{lstlisting}

The stream properties are optional in Tydi-lang. For properties not specified, the Tydi-lang compiler will use the default value. Table \ref{table:stream_properties} shows the default values and syntax for each stream property. The following code snippets show some examples of stream properties:

\begin{lstlisting}[language=c++]
type stream0 = Stream(Bit(4));
type stream1 = Stream(Bit(4), d=2, c=6); //different stream properties are separated by ","
type stream2 = Stream(Bit(4), c=6, d=2); //the order of properties is trivial.
\end{lstlisting}

\begin{table}[]
\centering
\caption{Stream properties and default values}
\label{table:stream_properties}
\resizebox{\textwidth}{!}{
\begin{tabular}{l|c|c|c|c}
\hline
Property & Value type & Default value & Possible value & Syntax \\ \hline \hline
dimension & integer & 0 & all non-negative integer & d =\textless{}Exp\textgreater{} \\ \hline
user type & logical type & Null & all non-stream logical types & u =\textless{}LogicalType\textgreater{} \\ \hline
throughput & float & 1 & all non-negative floating number & t =\textless{}Exp\textgreater{} \\ \hline
synchronicity & string & "Sync" & "Sync", "Flatten", "Desync", "FlatDesync" & s =\textless{}Exp\textgreater{} \\ \hline
complexity & integer & 7 & 1-7 & c=\textless{}Exp\textgreater{} \\ \hline
direction & string & "Forward" & "Forward", "Reverse" & r=\textless{}Exp\textgreater{} \\ \hline
keep & bool & false & true, false & x=\textless{}Exp\textgreater{} \\ \hline
\end{tabular}
}
\end{table}

The code files of each component are provided below:
\begin{itemize}
    \item Type alias system: \href{https://github.com/twoentartian/tydi-lang/blob/main/tydi_lang_raw_ast/src/variable.rs}{tydi-lang/tydi\_lang\_raw\_ast/src/variable.rs}
    \item \texttt{Bit} and \texttt{Null} type: \href{https://github.com/twoentartian/tydi-lang/blob/main/tydi_lang_raw_ast/src/bit_null_type.rs}{tydi-lang/tydi\_lang\_raw\_ast/src/bit\_null\_type.rs}
    \item \texttt{Group} and \texttt{Union} type: 
    \href{https://github.com/twoentartian/tydi-lang/blob/main/tydi_lang_raw_ast/src/group_union_type.rs}{tydi-lang/tydi\_lang\_raw\_ast/src/group\_union\_type.rs}
    \item \texttt{Stream} type:
    \href{https://github.com/twoentartian/tydi-lang/blob/main/tydi_lang_raw_ast/src/steam_type.rs}{tydi-lang/tydi\_lang\_raw\_ast/src/steam\_type.rs}
    \item Evaluating logic types:
    \href{https://github.com/twoentartian/tydi-lang/blob/main/tydi_lang_parser/src/evaluation_type.rs}{tydi-lang/tydi\_lang\_parser/src/evaluation\_type.rs}
\end{itemize}

Please note that some syntax combinations can also pass the parser check and pass the evaluation but do not guarantee correctness. For example, the following code can be compiled correctly:

\begin{lstlisting}[language=c++]
package tpch;

type byte = Bit(8); 
type rgb_alias = Group rgb {  //define an alias of Group RGB while declaring it
  const x = 8,
  r: Bit(x),
  g: Bit(x),
  b: Bit(x),
};

type rgb_stream = Stream(rgb_alias);
\end{lstlisting}

The code above makes an alias of logical type \texttt{rgb}. The code can be compiled correctly if only \texttt{rgb\_alias} is used in this code file. However, the logical type \texttt{rgb} is invisible to the package scope, and there are no scope relationships among them. So following code will result in errors because variable \texttt{x} is not found.

\begin{lstlisting}[language=c++]
package tpch;

const x = 8,
type rgb_alias = Group rgb {  //define an alias of Group RGB while declaring it
  r: Bit(x), // variable "x" not found because the there is no scope relationship with external package scope for rgb scope
  g: Bit(x),
  b: Bit(x),
};

type rgb_stream = Stream(rgb_alias);
\end{lstlisting}

My recommendation for the above issue is to only use the standard syntax at the beginning of this section to ensure correctness.

\subsection{Package and cross-package reference}
\label{subsubsection:Package}

Each Tydi-lang source file will be treated as an isolated package in the compiler. The package name must be declared at the first statement of the file (empty space does not count). An "import" statement must be declared in the package scope (directly in the file scope, cannot be in any other scopes such as group scope or streamlet scope) to access the language elements in other source files. The package name must be identical to the package name in the imported source file. After importing the external package, all language elements in that package scope will be accessible to this file. Use following syntax \texttt{\{PackageName\}.{ID}} to access language elements in package whose name is \texttt{PackageName}. The following code snippet illustrates accessing variables in file "simple\_1.td" from file "simple\_0.td".

The content for simple\_0.td:
\begin{lstlisting}[language=c++]
package simple_0;
import simple_1;

const i1: int = 1 + 100;
const external_var0 = simple_1.i1 + 10;             //access external variables
const external_flag0 = false || simple_1.flag;      //access external variables
\end{lstlisting}

The content for simple\_1.td:
\begin{lstlisting}[language=c++]
package simple_1;
const i1 = 100;
const flag = true;
const i2 = 500;
\end{lstlisting}

Please notice that the evaluation of cross-package variables also follows the rule of "lazy evaluation" (mentioned in Section \ref{subsubsection:lazy_evaluation}). A brief explanation of "lazy evaluation" is that the compiler only evaluates the value required by other values, and unused variables will not be evaluated. In the above code example, if we evaluate all variables in "simple\_0.td", the variable "i2" in "simple\_1.td" will not be evaluated. The corresponding code structure representation is provided below (the format of code structure representation is explained in Section \ref{subsubsection:evaluated_code_structure}):

\begin{lstlisting}[language=c++]
Project(test_project){
  Package(simple_0){
    Scope(package_simple_0){
      Variables{
        $package$simple_1:PackageType(NotInferred(""))
        external_flag0:bool(true)
        external_var0:int(110)
        i1:int(101)
        $package$simple_0:PackageType(NotInferred(""))
      }
    }
  }
  Package(simple_1){
    Scope(package_simple_1){
      Variables{
        i1:int(100)
        $package$simple_1:PackageType(NotInferred(""))
        flag:bool(true)
        i2:UnknownType(NotInferred("500"))
      }
    }
  }
}
\end{lstlisting}

When the compiler finds \texttt{"package simple\_0"} and \texttt{"import simple\_1"} in the source file, two magic variables will be created in the package scope: \texttt{"\$package\$simple\_1"} and \texttt{"\$package\$simple\_0"}. Developers will never define a variable with the same name because the variable names contain "\$", which is invalid at the parser stage. When the compiler finds developers are using variables in another scope, it will first check the existence of the magic variable to see whether the developer imports the package. The compiler will visit the target package scope to evaluate the specified variables after checking the existence of the package variable and return errors if the target package does not exist in the project or the evaluation of external variables fails.

There are other "magic identifiers" in Tydi-lang to separate the usual variable ids and internal ids. These ids are located here:
\href{https://github.com/twoentartian/tydi-lang/blob/main/tydi_lang_parser/src/built_in_ids.rs}{tydi-lang/tydi\_lang\_parser/src/built\_in\_ids.rs}

The magic variable of the self package enables the possibility of using the package level variables rather than the nearest variable according to the scope relationship. This feature can solve the issue in some cases where the variables are shadowed by local variables. For example, the following code uses the package level variable \texttt{"i"} rather than the variable \texttt{"i"} in local scope.

\begin{lstlisting}[language=c++]
package tpch;

const i = 16;
type byte = Bit(8);

type Group rgb {
  const i = 8,
  r: Bit(tpch.i),   //r=Bit(16)
  g: Bit(i),        //g=Bit(8)
  b: Bit(i),
};
\end{lstlisting}

The source code to manage the project and packages are here: 
\href{https://github.com/twoentartian/tydi-lang/blob/main/tydi_lang_parser/src/evaluation.rs}{tydi-lang/tydi\_lang\_parser/src/evaluation.rs}
The implementation of evaluating cross-package variables, logical types, streamlets, and implementations are distributed to their evaluation code.

In addition, at the time of writing this thesis, Tydi-IR does not support cross-package references yet. This limitation means that cross-package references can be evaluated but cannot be generated to Tydi-IR. Because there are no variables in Tydi-IR, using cross-package variables is safe in the Tydi-lang compiler, but using other cross-package features will result in a generation error (but you can still get the evaluation result).

\subsection{Streamlet and port}
\label{subsubsection:Streamlet}
Streamlet describes the interface of a component. Its role is similar to the "entity" in VHDL, except that streamlets use Tydi typed interfaces, and the ports are bound to clockdomains. The following code snippet shows some examples of declaring streamlets and ports. Notice that there is a comma that separates statements in streamlet.

\begin{lstlisting}[language=c++]
package tpch;

type Group rgb {
  r: Bit(8),
  g: Bit(8),
  b: Bit(8),
};

#streamlet documentation#       //a documentation for streamlet
streamlet rgb_bypass {          //declaring a streamlet called rgb_bypass
  input: Stream(rgb) in,        //{port name} : {Logical type} in/out,
  output: Stream(rgb) out,
};

type rgb_stream = Stream(rgb);
const cd:clockdomain = "any string";

#streamlet documentation#
streamlet rgb_bypass2 {
  input: rgb_stream in `cd,     //optional clockdomain, starts with "`" (back single quote), followed by clockdomain variable name
  output: rgb_stream out `cd,
};
\end{lstlisting}

The streamlet documentation is a sentence wrapped by two "\#" to explain the high-level meaning of this streamlet. The documentation will be transformed into low-level VHDL documentation. This feature can increase productivity and allow better cooperation among high-level and low-level developers. The implementation also supports the documentation system and will not be discussed again in the implementation section. 

Each port in the streamlet must have the following properties: port name, logical type, port direction, and clockdomain. Port names cannot be identical in the same streamlet. The clockdomain is an optional value. It will resolve to the default clockdomain expression provided by the Tydi-lang compiler if the developers do not provide one.

The streamlet scope also supports defining variables and logical types to record some properties. Developers can use \texttt{"streamlet"} keyword to extract the variable value outside the streamlet scope. For example, the following code snippet shows defining variables and extracting them.

\begin{lstlisting}[language=c++]
package tpch;

...

streamlet rgb_bypass2 {
  const delay = 10,             //defining a variable
  type t = Bit(8),              //define a logical type

  input: rgb_stream in `cd,
  output: rgb_stream out `cd,
};

const delay = streamlet rgb_bypass2.delay;  //extra the delay variable in "rgb_bypass2", delay = 10

\end{lstlisting}

This feature can be used to pass specifications/configurations from one component to another one. For example, generate a new clockdomain and perform design verification with the help of assertion in Section \ref{subsubsection:Assertion}.

The source code location is provided in the following code:
\begin{itemize}
    \item Defining streamlet concept:
    \href{https://github.com/twoentartian/tydi-lang/blob/main/tydi_lang_raw_ast/src/streamlet.rs}{tydi-lang/tydi\_lang\_raw\_as/src/streamlet.rs}
    \item Evaluating streamlet (including evaluating streamlet template):
    \href{https://github.com/twoentartian/tydi-lang/blob/main/tydi_lang_parser/src/evaluation_streamlet.rs}{tydi-lang/tydi\_lang\_parser/src/evalua-tion\_streamlet.rs}
\end{itemize}

\subsection{Implementation, instance and connection}
\label{subsubsection:Implement}
Implementation describes the structure of a component by characterizing its internal instances and connections. The term "instances" refers to an instance of another implementation. The term "connection" means connecting two ports. The two ports of a valid connection must have compatible logical types, opposite port directions, and the same clockdomain. The following code snippet shows an implementation. Notice that statements in implementation are separated by a comma.

\begin{lstlisting}[language=c++]
package tpch;

type Group rgb {
  r: Bit(8),
  g: Bit(8),
  b: Bit(8),
};
type rgb_stream = Stream(rgb);

streamlet rgb_bypass {
  input: rgb_stream in,
  output: rgb_stream out,
};

#implement documentation#
impl impl_rgb_bypass of rgb_bypass {    //declare an implementation called "impl_rgb_bypass" and its interface (streamlet) is "rgb_bypass"
  input => output,                      //connect the input port and the output port.
};
\end{lstlisting}

There are two methods to determine whether two logical types are compatible. The first method is called "strict type checking" which checks whether the two type variables resolve to the same logical type value. The code example above uses this method because both ports use \texttt{"rgb\_stream"} as the logical type variable. The second method is called "compatible type checking" which compares the content of the logical types. For example, two logical variables that both are defined as \texttt{Bit(8)} are compatible. The first method is the default method in the Tydi-lang compiler, and developers do not need to declare it explicitly. The second method requires users explicitly add \texttt{"@NoStrictType@"} at the end of connections. Connecting two compatible ports without \texttt{"@NoStrictTye@"} results in a warning in the DRC stage. The following code illustrates the second type of checking method.

\begin{lstlisting}[language=c++]
package tpch;

type Group rgb {
  r: Bit(8),
  g: Bit(8),
  b: Bit(8),
};
type rgb_stream = Stream(rgb);

streamlet rgb_bypass2 {
  input: Stream(rgb) in,    //Stream(rgb) is a logical type
  output: Stream(rgb) out,  //Stream(rgb) is a new logical type, the content is same as the previous one
};

impl impl_rgb_bypass2 of rgb_bypass2 {
  input => output @NoStrictType@,   //explicitly add @NoStrictType@
};

impl impl_rgb_bypass3 of rgb_bypass2 {
  input => output,   //result in a warning in DRC
};
\end{lstlisting}

It is possible to declare a FIFO buffer in the connection, whose buffer size can be stated as an integer expression. It is also possible to specify the name of a connection. The Tydi-lang compiler will automatically generate a name with its starting token position and end token position to generate a connection name for connections with unspecified names. The following code snippet shows an example of a FIFO and specifies a connection name.

\begin{lstlisting}[language=c++]
impl impl_rgb_bypass of rgb_bypass {
  input =1=> output "input2output" @NoStrictType@,        //the size of FIFO buffer is 1, the connection name is "input2output"
};
\end{lstlisting}

The syntax to define an instance inside an implementation is similar to the syntax of defining an instance of a class in C++. The following code snippet shows a basic example.

\begin{lstlisting}[language=c++]
streamlet rgb_bypass {
  input: rgb_stream in,
  output: rgb_stream out,
};

impl impl_rgb_bypass_inner of rgb_bypass {
  input => output,
};

impl impl_rgb_bypass of rgb_bypass {
  instance inner(impl_rgb_bypass_inner),    //declare an instance of "impl_rgb_bypass_inner", the instance name is "inner"
  input => inner.input, //"inner.input" refers to the "input" port on instance "inner"
  inner.output => output,
};
\end{lstlisting}

For streaming components,  a port without a connection is not allowed because this port will be blocked due to the handshaking mechanism in Tydi-spec. Tydi-IR also checks that each port has a valid connection. However, this is no restriction in Tydi-lang because sugaring (mentioned in Section \ref{subsubsection:sugaring}) will add connections and corresponding instances automatically.

For components that can only be described with low-level HDL, Tydi-lang provides a keyword called "external" to specify that this implementation should have an empty implementation body and tell the compiler to find the implementation on the lower-level side. The following code snippet shows an example of an external implementation.

\begin{lstlisting}[language=c++]
streamlet duplicator_s {
  input: data_type in,
  output: data_type [output_channel] out,
};

external impl duplicator_i of duplicator_s {

};
\end{lstlisting}

\subsection{Array}
\label{subsubsection:Array}
The term "array" in Tydi-lang means grouping several similar targets with a single name and using an index to access every target. The array concept can be applied to basic variables, ports, and instances. Arrays of basic variables support "+" operator to insert variables at the beginning/end of the array. Notice that the type of the inserted value must be identical to the element type. The following code snippet shows some examples of declaring basic variable arrays and operations on them.

\begin{lstlisting}[language=c++]
const array_exp_0 = {1,2,3,4,5};        //array of integer
const array_exp_1 = {true,true,false};  //array of boolean
const array_exp_2 = {"123", "456"};     //array of string
const array_exp_3 = {1.1,2.1,3.1};      //array of float
const array_exp_4 = array_exp_3 + 50.5; //append a float to a float array
const array_exp_5 = 50.5 + array_exp_3; //insert a float at the beginning of a bool array
const array_exp_6 = true + {true,false};//insert a bool at the beginning of a bool array
\end{lstlisting}

Port and instance can also be declared as port arrays. The following code snippet shows port arrays and instance arrays.

\begin{lstlisting}[language=c++]
streamlet data_bypass_channel {
  inputs: bit8_stream [channel] in `"10kHz",    //declaring a port array
  outputs: bit8_stream [channel] out `"10kHz",  //declaring a port array
};

impl impl_data_bypass_channel of data_bypass_channel {
  instance bypass(impl_data_bypass) [channel],  //declaring an instance array
  
  ...
  
};
\end{lstlisting}

The syntax to access an element in an array is \texttt{"ArrayName[Index]"}. All places that can be a variable can also be an array element. Tydi-lang does not support two or higher dimensional arrays because the current compiler implementation is at the prototype level. Higher-dimensional arrays on the software side can be flattened to 1-d arrays in Tydi-lang with the price of losing readability.  

In Tydi-lang, the "for" block can be used to iterate an array or generate connections/instances from a basic variable array. The elaboration of the "for" block will be in Section \ref{subsubsection:If_and_for_block}.

\subsection{If and for block}
\label{subsubsection:If_and_for_block}
The "if" and "for" blocks automatically generate instances and connections in implementations. Their syntax is similar to the "if" and "for" syntax in modern software programming languages, such as C++ and Python. The following code snippet illustrates a code sample where the implementation "impl\_data\_bypass\_channel" can automatically generate inner structure according to two variables, "use\_data\_bypass2" and "channel".

\begin{lstlisting}[language=c++]
package main;

type bit8_stream = Stream(Bit(8), d = 5, t = 2.5);

//define impl_data_bypass
streamlet data_bypass {         
  input: bit8_stream in,
  output: bit8_stream out,
};
impl impl_data_bypass of data_bypass {
  input => output,
};

//define impl_data_bypass2
streamlet data_bypass2 {
  input: bit8_stream in,
  output: bit8_stream out,
};
impl impl_data_bypass2 of data_bypass2 {
  input => output,
};

const channel = 10;                 //control the channel count
streamlet data_bypass_channel {
  inputs: bit8_stream [channel] in `"10kHz",
  outputs: bit8_stream [channel] out `"10kHz",
};

const use_data_bypass2 = true;      //this variable can control which implementation to use
impl impl_data_bypass_channel of data_bypass_channel {
  if (use_data_bypass2) {
    instance bypass(impl_data_bypass) [channel],
    for i in (0=1=>channel) {       //for block, 0=1=>channel is an sugar expression to generate an int array
      bypass[i].output => outputs[i],
      inputs[i] => bypass[i].input,
    }
  }
  //elif ({BoolVariable}) {}        //elif block is optional
  else {
    instance bypass(impl_data_bypass2) [channel],
    for i in (0=1=>channel) {
      bypass[i].output => outputs[i],
      inputs[i] => bypass[i].input,
    }
  }
};
\end{lstlisting}

For "if" syntax, the variable inside the brackets must be a boolean value. The content in the "if" scope will be copied to the outer scope if the evaluation result of the variable is true. Otherwise, nothing happens. The "elif" and "else" blocks are optional, and users need to provide a boolean variable in the "elif" bracket. A new variable will be created in the "for" scope, which is only accessible in the "for" scope. The variable name is stated as the identifier after the "for" keyword ("i" in the above example). The new variable type is the same as the array element type. The identifier after the "in" keyword must refer to an array of basic types. The content in the "for" scope will be copied to the external scope with each element value in that array.

The instance cannot be declared in the "for" scope because there will be multiple instances with the same name after copying to the external scope. Appendix \ref{subsubsection:Duplicated_identifier_issue_in_for_if_expansion} records this issue details and proposed solutions.

\subsection{Template}
\label{subsubsection:Template}
The template in Tydi-lang is similar to the template system in Rust and C++. "Template" means this is not a specific streamlet or implementation but rather a process to generate a series of streamlets or implementations according to the template arguments. The template system is built based on the variable system and logical type system. An example of a template streamlet is available below:

\begin{lstlisting}[language=c++]
package main;

streamlet duplicator_s<data_type: type, output_channel: int> {      //a template streamlet
  input: data_type in,
  output: data_type [output_channel] out,
};
\end{lstlisting}

In the above snippet, "data\_type" and "output\_channel" are two template arguments. Tydi-lang supports using five basic variables and logical types as template arguments. When declaring template arguments, the corresponding template variables will be declared in the streamlet (or implementation) scope and can be directly used as expressions. For the above example, "data\_type" is used as the port type of the "input" and the "output" port. The "output\_channel" is used as the size of the "output" port array because it is declared as an integer. Multiple template arguments are separated by a comma. 

Tydi-lang supports passing the template arguments of an implementation to a streamlet. This pattern can be illustrated with the following example.

\begin{lstlisting}[language=c++]
package main;

type bit8_stream = Stream(Bit(8), d = 5, t = 2.5);

const eight = 8;
type Group rgb {
  r: Bit(eight),
  g: Bit(eight),
  b: Bit(eight),
};
type rgb_stream = Stream(rgb);

streamlet data_bypass<data_type: type> {
  input: data_type in,
  output: data_type out,
};
impl impl_data_bypass<data_type: type> of data_bypass<type data_type> { //passing template arguments from implementation to streamlet
  input => output,
};

streamlet data_demux<channel:int, data_type: type, cd:clockdomain> {    //use clockdomain as template argument
  inputs: data_type [channel] in `cd,
  outputs: data_type [channel] out `cd,
};
impl impl_data_demux<channel:int, data_type: type, cd:clockdomain> of data_demux<channel, type data_type, cd> { //data_type is a logical type, so we must add a "type" keyword before it.
  instance bypass(impl_data_bypass<type data_type>) [channel],
  for i in (0=1=>channel) {
    bypass[i].output => outputs[i],
    inputs[i] => bypass[i].input,
  }
};

const cd0: clockdomain = "100MHz";
const cd1: clockdomain;
impl data_demux_bit8_5(impl_data_demux<5, type bit8_stream, cd0>);  //declare implementations based on implementation templates
impl data_demux_rgb_100(impl_data_demux<20, type rgb_stream, cd1>); 
\end{lstlisting}

In line 17 and line 25, the code defines an implementation template based on an instance of a streamlet template. The instance syntax is \texttt{"TemplateName\textless{}TemplateArgExp\textgreater{}"}. Multiple template argument expressions are separated by a comma. Notice that a "type" keyword must be put before the template argument expression if it is a logical type. This "type" keyword will tell the compiler to find the identifier in logical type scopes rather than in variable scopes. Similarly, in line 26, the code declares an instance array based on the implementation template.

Streamlet templates and implementation templates will not be compiled to Tydi-IR because they are not describing any components. In line 35 and line 36, two implementations are declared based on the templates, and the two implementations will be compiled to Tydi-IR. The syntax to declare implementations based on implementation templates is similar to declaring instances in implementation.
 
The Tydi-lang compiler will never evaluate a template itself until it is instantiated. The first step of evaluating a template is copying the template's content to the instantiation. Then the compiler assigns the template argument expressions to the corresponding template variables. Finally, evaluate the copied version of the template.

The implementation templates in the code structure have template arguments starting with a "@" while normal implementations never have. The following code snippets shows the difference. 

\begin{lstlisting}[language=c++]
//This is a normal implementation called "orders_i"
Implement(orders_i)<NormalImplement> -> Streamlet(orders_s){
  Scope(implement_orders_i){
    ScopeRelations{
      --ImplementScope-->package_tpch
    }
  }
  simulation_process{None}
}

//This is a implementation template, it has one template argument, which is marked as @LogicalDataType(DummyLogicalData). The interface streamlet is also a template instance so it's not known yet. The name of the streamlet template is "void_s" and the template argument expression is "type_in". "type_in" is the template argument of "void_i" and it exists in the implementation scope as a result of template argument. 
Implement(void_i)<@LogicalDataType(DummyLogicalData)> -> ProxyStreamlet(void_s<@type_in>){
  Scope(implement_void_i){
    Variables{
      type_in:DummyLogicalData(NotInferred("$arg$type_in"))
    }
    ScopeRelations{
      --ImplementScope-->package_tpch
    }
  }
  simulation_process{None}
}

//This is an instance of the above implement template, please notice that the implementation name is replaced by "void_i@Stream(SQL_char1_stream)", this is an invalid identifier in Tydi-lang so the names of template instances will never be the same as the developers' identifiers. The streamlet name is also a generated name. The template argument "type_in" is also replaced by the template argument expression.
Implement(void_i@Stream(SQL_char1_stream))<NormalImplement> -> Streamlet(void_s@Stream(SQL_char1_stream)){
  Scope(implement_void_i@Stream(SQL_char1_stream)){
    Types{
      type_in:Stream(SQL_char1_stream){
        DataType=Bit(8)
        dimension=1, user=DataNull, throughput=1, synchronicity=Sync, complexity=7, direction=Forward, keep=false
      }
    }
    ScopeRelations{
      --ImplementScope-->package_tpch
    }
  }
  simulation_process{None}
}

//this a normal streamlet called "orders_s"
Streamlet(orders_s)<NormalStreamlet>{
  Scope(streamlet_orders_s){
    ScopeRelations{
      --StreamletScope-->package_tpch
    }
    Ports{
      o_custkey:Port(Stream(int_stream),out) `DefaultClockDomain
      o_orderdate:Port(Stream(date_stream),out) `DefaultClockDomain
      o_totalprice:Port(Stream(SQL_decimal_15_2_stream),out) `DefaultClockDomain
      o_shippriority:Port(Stream(int_stream),out) `DefaultClockDomain
      o_comment:Port(Stream(varchar_stream),out) `DefaultClockDomain
      o_clerk:Port(Stream(SQL_char15_stream),out) `DefaultClockDomain
      o_orderkey:Port(Stream(int_stream),in) `DefaultClockDomain
      o_orderstatus:Port(Stream(SQL_char1_stream),out) `DefaultClockDomain
      o_orderpriority:Port(Stream(SQL_char15_stream),out) `DefaultClockDomain
    }
  }
}

//This is a streamlet template, similar to implementation template, "type_in" is a template argument.
Streamlet(void_s)<@LogicalDataType(DummyLogicalData)>{
  Scope(streamlet_void_s){
    Variables{
      type_in:DummyLogicalData(NotInferred("$arg$type_in"))
    }
    ScopeRelations{
      --StreamletScope-->package_tpch
    }
    Ports{
      input:Port(VarType(type_in),in) `DefaultClockDomain
    }
  }
}
        
//this is an instance of a streamlet template, the generated name is "void_s@Stream(SQL_decimal_15_2_stream)"
Streamlet(void_s@Stream(SQL_decimal_15_2_stream))<NormalStreamlet>{
  Scope(streamlet_void_s@Stream(SQL_decimal_15_2_stream)){
    Types{
      type_in:Stream(SQL_decimal_15_2_stream){
        DataType=DataGroup(SQL_decimal_15_2)
        dimension=1, user=DataNull, throughput=1, synchronicity=Sync, complexity=7, direction=Forward, keep=false
      }
    }
    ScopeRelations{
      --StreamletScope-->package_tpch
    }
    Ports{
      input:Port(Stream(SQL_decimal_15_2_stream),in) `DefaultClockDomain
    }
  }
}
\end{lstlisting}

The Rust source files of implementing template are distributed files of each component. For example, streamlet template is implemented with streamlet.

\subsection{Use components as template arguments}
\label{subsubsection:use_components_as_templates}

In some cases, users might want to describe components with known interfaces but with unknown implementations. Suppose we have an adder A whose delay is two cycles and does not support pipeline, but the desired input data rate is one addition per cycle. The solution is to use a data multiplexer and a data demultiplexer to split the data into two adders. In the future, low-level developers might design adder B with a 4-clock delay but much less area. Developers need to manually redesign the multiplexer and demultiplexer to meet the data rate requirement. For such a case, we can set the adder as a template component and expose its interface to the demultiplexer and multiplexer. The following code snippet shows an example of using a template component.

\begin{lstlisting}[language=c++]
package main;

type Group rgb {
  r: Bit(eight),
  g: Bit(eight),
  b: Bit(eight),
};

type rgb_stream = Stream(rgb);

//we define a streamlet called "component"
streamlet component {
  input: rgb_stream in,
  output: rgb_stream out,
};

//define three implementations of "component", here for simplicity the three implementations are the same
impl component_impl0 of component {
  input => output,
};

impl component_impl1 of component {
  input => output,
};

impl component_impl2 of component {
  input => output,
};

//an example of using abstract implement
streamlet larger_component {
  input: rgb_stream [2] in,
  output: rgb_stream [2] out,
};

impl impl_larger_component<ts: impl of component> of larger_component { //"component" is a streamlet name, notice that the keyword "impl of" before the streamlet name
  instance inst(ts) [2],
  for i in (0=1=>2) {
    input[i] => inst[i].input,
    inst[i].output => output[i],
  }
};

impl impl_larger_component0(impl_larger_component<impl component_impl0>);   //use an implementation of "component" streamlet to instantiate the template, notice that the keyword "impl" before the implementation name.
impl impl_larger_component1(impl_larger_component<impl component_impl1>);
\end{lstlisting}

In the above example, "impl\_larger\_component" is a template that receives an implementation as an argument. The interface of the argument is specified as "component". The identifier after the "impl of" keyword must be a streamlet. An implementation of that streamlet must be provided to instantiate the template, as shown in lines 44 and 45.

\subsection{Assertion}
\label{subsubsection:Assertion}
As mentioned previously, Tydi-lang is an abstract hardware description language where developers can use constant variables and logical types to describe hardware components. The assertion is designed to set limitations for the abstract hardware by restricting the variable values. For example, the following code illustrates setting a limitation for the template arguments. 

\begin{lstlisting}[language=c++]
package main;
type Group rgb {
  const x = 8,
  r: Bit(x),
  g: Bit(x),
  b: Bit(x),
};
streamlet component<data:type> {
  const x = type data.x,
  assert(x == 8),           //assert x == 8
  input: Stream(data) in,
  output: Stream(data) out,
};
impl component_impl<data:type> of component<type data> {
  input => output,
};
impl component_impl0(component_impl<type rgb>);     //use logical type rgb as template argument
\end{lstlisting}

The assertion is a built-in function that uses "assert" as the identifier and receives one argument with boolean type. The built-in function is a powerful system that can do various processing on language elements, such as transform logical types and decay logical types back to variable values. However, due to time limitations, the current Tydi-lang compiler only supports the assertion function. I put three of my personal proposed builtin functions in Table \ref{table:proposed_builtin_functions} (Appendix) for future Tydi-lang developers' reference. The functions in this proposal can be applied to support assertions on logical types.

%% file: 04-tydi-language-front-end.tex
\chapter{Tydi language compiler frontend}
\label{subsubsection:tydi_lang_front_end}

\section{Introduction to Tydi-lang frontend}
\label{subsecion:introduction_to_tydi_lang_frontend}
The previous chapter illustrates the Tydi-lang specifications and syntax, which should be materialized as a compiler to automatically compile the source code that meets the specification and syntax to a lower-level representation. Because Tydi-IR already provides a direct representation of Tydi-spec, compiling Tydi-lang to Tydi-IR becomes a convenient way to support various backends. Meanwhile, compiling to Tydi-IR also provides flexibility to cooperate with other future frontends. For example, multiple frontends generate Tydi-IR, which could mutually access components generated from other frontends.

The structure of the Tydi-lang frontend is similar to that of a software compiler frontend. For example, both have parsers, name resolutions, references, and scopes. The logical type system in Tydi-lang is also similar to the user-defined type system in general software languages. The difference is that the evaluation of all variables is performed while compiling the Tydi-lang source code. In contrast, the evaluation of variables is usually performed during the runtime for most software programming languages. Some hardware-specific properties also cause some differences, such as the design rule check, hardware simulation, and generating testbench.

This chapter focuses on elaborating on the compiling process of the Tydi-lang frontend and some possible optimizations, as well as some intermediate representations that can be used for debugging the Tydi-lang source code or continuing developing Tydi-lang infrastructures. 

\section{Overall work flow}
\label{subsubsection:overall_work_flow}

\begin{figure}
    \centering
     \includegraphics[width=\textwidth]{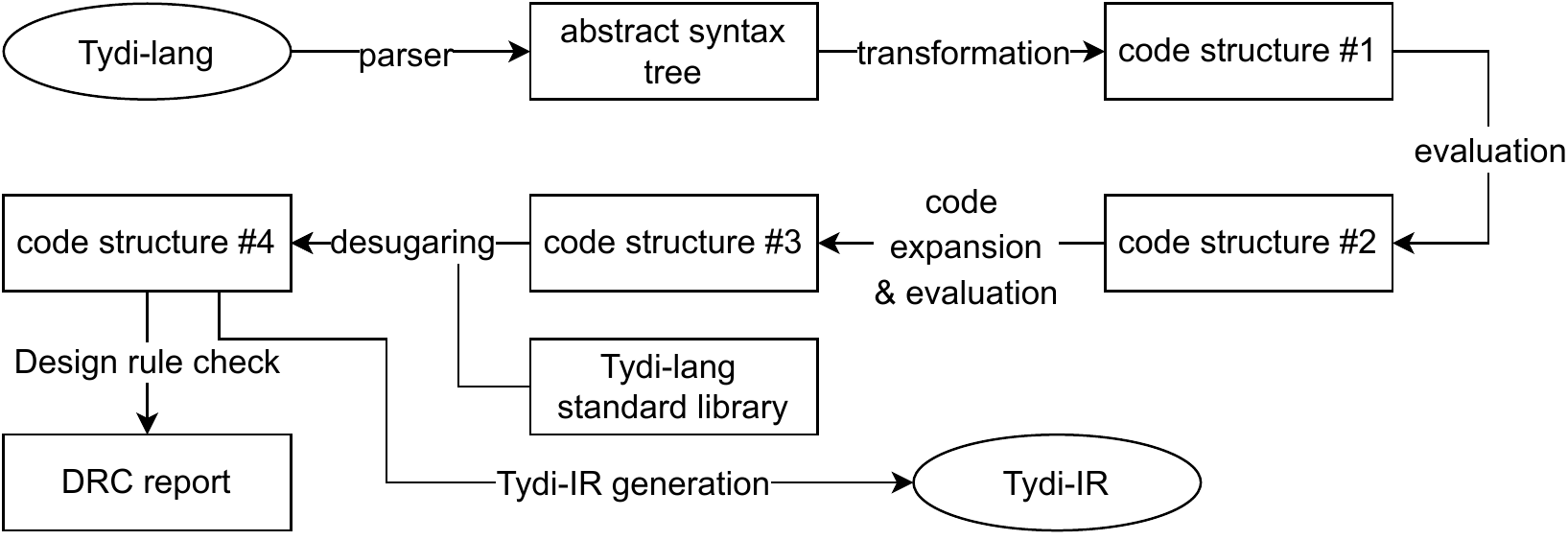}
    \caption{Overview of the Tydi-lang frontend}
    \label{fig:tydi_lang_frontend_work_flow}
\end{figure}

This section provides an overview of the Tydi-lang compiler frontend. As mentioned in Figure \ref{fig:tydi_lang_tool_chain_work_flow}, the frontend compiles the Tydi-lang source code to Tydi-IR. Figure \ref{fig:tydi_lang_frontend_work_flow} shows the detailed steps of compiling from Tydi-lang to Tydi-IR. A parser called PEST \cite{pest} transforms the plain-text Tydi-lang source code to a tree structure, known as abstract syntax tree (AST). The AST can be transformed to the "code structure", a memory structure which contains extra variables such as evaluation flags, multi-threading locks, name resolution result, etc. Each step in the Tydi-lang frontend creates a new "code structure". So the number after the "code structure" in Figure \ref{fig:tydi_lang_frontend_work_flow} indicates the version. Tydi-lang provides some sugaring syntax for developers, so the Tydi-lang compiler needs to perform de-sugaring, which requires some components in the Tydi-lang standard library. The Tydi-lang standard library collects many fundamental and useful component templates that can assist in developing hardware. The design rule check (DRC) is designed to identify errors on the Tydi-lang level. Some high-level errors might hide themselves in after generating low-level representations. For example, two ports with different but same-width logical types are not compatible and should not be connected together. After generating the low-level representation, they become compatible because they have the same bit width. The DRC in Tydi-lang level can identify these high-level errors in advance. 

All output files mentioned in this chapter are available in the compiler output folder. A sample output folder might have following structure.

\href{https://github.com/twoentartian/tydi-lang/tree/main/CookBook/12_tpch_sql3/build}{tydi-lang/CookBook/12\_tpch\_sql3/build}

The \texttt{"0\_ast"} folder contains the AST tree for all source files. Each file corresponds to a Tydi-lang source file. The \texttt{"1\_parser\_output.txt"} file records the "code structure \#1". The \texttt{"2\_evaluation\_output.txt"} file records the code structure after evaluation and expansion, corresponding to "code structure \#3". The \texttt{"2\_evaluation\_output\_after\_sugaring.txt"} represents the "code structure \#4". The compiler will generate a DRC report if the DRC flag is set. An error report will be generated if the compiling fails, and the output files before the error occurs will be generated as usual. The \texttt{"3\_til"} folder contain the generated Tydi-IR and the \texttt{"4\_vhdl"} folder contains the final output VHDL files.

\section{Mutable memory structure in Rust}
\label{subsubsection:Mutable_memory_structure_in_Rust}
This section explains the mutable/immutable reference issue introduced by Rust. This issue is extremely important, and failing to deal with this issue will rapidly increase the difficulty of designing the Tydi-lang compiler. For future developers that continue working on Tydi and plan to use Rust as the developing language, please read this section carefully and investigate all possible solutions before you really write any code for Tydi.

In Rust, all variables must be declared as mutable or immutable. This is a common feature in many other languages such as C++ and Java. The reference of the variable (equivalent to the pointer in C++) must also be declared as mutable or immutable, and each reference has its own lifetime \cite{rust-lifetime}. The lifetime means the code region from where it is declared to where it is last used. Immutable variables can only have immutable references, and mutable variables can have multiple immutable references or one mutable reference. Notice that having multiple mutable references with overlapped lifetime for a single mutable variable is not allowed. One mutable reference and one immutable reference with overlapped lifetime are also not allowed. The check of lifetime overlap is performed at the compile stage, and this mechanism ensures data consistency in multi-threading environments. 

However, the mutable reference and lifetime mechanism causes big troubles for developers who want to design their own data structures. For example, developers who design a tree structure might find that the tree is completely immutable even though each node is mutable, because other nodes hold a mutable reference of that node, and that mutable reference prevents making modifications to that node. 

To solve the issue, some students in our group (accelerated big data group) use Salsa \cite{rust-salsa}. Salsa can store the values and functions in a key-based in-memory database. Constructing a tree is equivalent to adding more entries to the database. The difficulty is that you need to construct the tree from the leaf nodes because the value becomes immutable once put into database. In other words, you cannot add more leaf nodes if the parent nodes are stored into database. The Tydi-spec prototype \cite{tydi-prototype} and the Tydi backend \cite{tydi-backend} use this method. This problem for this solution is the loss in flexibility. The data stored in the database is still immutable and users must finalize the value before storing it. In some cases, for example, referring a value in another file which is not analyzed yet, storing the finalized value in Salsa becomes a quite expensive operation because analyzing another file introduces more dependencies. Because compilers cannot predict the user input, more complicated dependencies, such as mutual references between files, requires much more design effort.

Another solution to solve the immutable/mutable reference issue is using the unsafe Rust feature. Unsafe Rust removes the lifetime checking mechanism in safe Rust. The Rust community recommends this method for implementing complex data structures, which is exactly our need. However, users need to manage the data consistency by themselves in the unsafe Rust and write safe-Rust interfaces to operate on unsafe-Rust data. The Tydi-lang compiler does not use this method because using unsafe Rust might introduce more potential risks, which is unacceptable for a nine-month thesis.

The last solution is using the read/write lock (RwLock) provided in Rust standard library. The RwLock provides a way to obtain a mutable reference at any time, but only one mutable reference can exist at any time point. The thread that tries to get the second mutable reference will be blocked until the first mutable reference runs out of its lifetime. In the previous tree example, each node will be stored in a RwLock, and other nodes hold an immutable reference of the lock. A mutable reference of the node can be obtained from the immutable reference of the lock when developers want to change its content. However, wrapping the data with locks introduces extra performance overhead. I did not measure the overhead yet because other choices are too complicated or risky. In addition, using locks always introduces deadlock problems. Developers should manage these locks carefully to avoid deadlocks.

We summarize the three solutions here to assist future Tydi developers to make design decisions.

\begin{itemize}
    \item Using Salsa: \newline \textbf{Pros}: it is a kind of in-memory database; there is not too much performance overhead(Salsa calculates hash for values, which might be the only performance overhead); many examples in Tydi-lang backend. \newline \textbf{Cons}: the data stored in Salsa database is immutable, and developers might need to insert new data to replace the old one to update the value.
    
    \item Using unsafe Rust: \newline \textbf{Pros}: many documents and examples on Rust official site; no performance overhead (at least in theory); the recommend way from Rust community. \newline \textbf{Cons}: developers need to ensure the data consistency in multi-threading cases by themselves.

    \item Using RwLock: \newline \textbf{Pros}: it is easy to use; examples available in Tydi-lang frontend. \newline \textbf{Cons}: the performance overhead might be large; developers must avoid deadlock in some cases.
\end{itemize}

\section{Parsing}
\label{subsubsection:Parsing}
This section illustrates the process of parsing the Tydi-lang source code to the abstract syntax tree (AST) and transforming AST to the "code structure \#1". The parser is called PEST \cite{pest}, and its working process can be briefly described as using a grammar file to define grammar rules. The PEST parser can automatically parse the source code to an AST according to the grammar file. The PEST grammar syntax will not be discussed here because this is not a major contribution. 
The PEST grammar file for Tydi-lang is located in the following link: 

\href{https://github.com/twoentartian/tydi-lang/blob/main/tydi_lang_parser/src/tydi_lang_syntax.pest}{tydi-lang/tydi\_lang\_parser/src/tydi\_lang\_syntax.pest}

\subsection{Parsed code structure}
\label{subsubsection:parse_code_structure}
There is a text representation of an AST in Tydi-lang. Every source file has its own AST in the compiling output folder, and developers can debug the grammar rules with this file. For an example of the text representation, let's consider the following sample Tydi-lang code snippet.

\begin{lstlisting}[language=c++]
Union A {
  a : Bit(10),
  b : Stream(A, d=0, t="user type"),
  c : Stream(A, t="user type"),
  d : Stream(A, d=0),
  e : Stream(A),
}
\end{lstlisting}

The code snippet defines a logical union type. The following text shows the parsed AST of the above logical union type. Please notice that "[]" represents a new hierarchy level. \texttt{"LogicalUnionType"} is at the top level.

\begin{lstlisting}[language=c++]
[LogicalUnionType(0, 134, [ID(6, 7), SubItemItem(12, 24, [ID(12, 13), LogicalType(16, 23, [LogicalBitType(16, 23, [Exp(20, 22, [Term(20, 22, [IntExp(20, 22, [INT_RAW_NORAML(20, 22)])])])])])]), SubItemItem(27, 61, [ID(27, 28), LogicalType(31, 60, [LogicalStreamType(31, 60, [LogicalType(38, 39, [LogicalUserDefinedType(38, 39, [ID(38, 39)])]), StreamPropertyDimension(39, 44, [Exp(43, 44, [Term(43, 44, [IntExp(43, 44, [INT_RAW_NORAML(43, 44)])])])]), StreamPropertyThroughput(44, 59, [Exp(48, 59, [Term(48, 59, [StringExp(48, 59, [STR(48, 59)])])])])])])]), SubItemItem(64, 93, [ID(64, 65), LogicalType(68, 92, [LogicalStreamType(68, 92, [LogicalType(75, 76, [LogicalUserDefinedType(75, 76, [ID(75, 76)])]), StreamPropertyThroughput(76, 91, [Exp(80, 91, [Term(80, 91, [StringExp(80, 91, [STR(80, 91)])])])])])])]), SubItemItem(96, 115, [ID(96, 97), LogicalType(100, 114, [LogicalStreamType(100, 114, [LogicalType(107, 108, [LogicalUserDefinedType(107, 108, [ID(107, 108)])]), StreamPropertyDimension(108, 113, [Exp(112, 113, [Term(112, 113, [IntExp(112, 113, [INT_RAW_NORAML(112, 113)])])])])])])]), SubItemItem(118, 132, [ID(118, 119), LogicalType(122, 131, [LogicalStreamType(122, 131, [LogicalType(129, 130, [LogicalUserDefinedType(129, 130, [ID(129, 130)])])])])])])]
\end{lstlisting}

All numbers in the AST indicate the token (character) index in the source code, e.g. "0, 134" means starting from the first token to the token at 134. The AST is a tree structure, for above example, A \texttt{LogicalUnionType} is composed of an identifier and several \texttt{SubItemItem}, each \texttt{SubItemItem} is made up of an identifier and a \texttt{LogicalType}. The elements in the AST will be transformed into a code structure. The transformation includes classifying the elements into different categories, such as logical types and streamlets, and adding invisible language elements, such as scope relationships. Examples of code structure will be provided in Section \ref{subsection:value_and_target_evaluation}.

\subsection{Multi-thread and multi-file parsing}
\label{subsubsection:multi_thread_multi_file_parsing}
Multi-file compiling is important for a large project containing many different source files. Parsing these source files can be optimized with multi-threading. Since there is no dependency issue among source files at the parser stage, the parsing is intrinsically parallelizable (The cross-file name resolution is performed at the evaluation stage).

Here I would like to mention the mutable/immutable memory issue in Rust here because, with immutable memory, the compiler needs to perform name resolution when seeing an identifier for the first time (might be at the parser stage). However, the situation would be complicated if the identifier is defined in another file that is not yet analyzed. Turning to analyze that file is not a solution because the dependency path might be extremely long. The above arguments are the major reasons that I choose to use RwLock to get mutable memory in Rust.

\subsection{Limitations for PEST}
\label{subsubsection:limitations_for_pest}
PEST is a lightweight parser, and some features common in the compiler parser area are not available in PEST. The most important missing feature is that PEST does not support left-recursive parsing. Left-recursive parsing is a case where a term starts with the term itself. For example, consider the following PEST grammar.

\begin{lstlisting}[language=c++]
Exp = { Exp ~ "+" ~ Exp }
\end{lstlisting}

This grammar is valid in many compiler workbenches such as Spoofax \cite{spoofax}. However, PEST does not accept this kind of grammar because parsing \texttt{Exp} will immediately result in parsing \texttt{Exp} again, causing infinite recursive parsing.

The math system in Tydi-lang contains many grammar rules similar to the above example. So I separate the meaning of \texttt{Exp} to two grammar syntax: \texttt{Exp} and \texttt{Term}. \texttt{Exp} is a math expression contains numbers and digits, while \texttt{Term} only includes numbers of bracket expression. Appendix \ref{appendix:wrong_precedence_for_unary_operator} also mentioned this issue.

The missing support of left-recursive parsing also causes the missing support of syntax-level precedence. For example, the expression \texttt{"1+5*9"} should be parsed with "5*9" first because it has higher precedence in math. PEST always parses source code from left to right, so the precedence information is lost on the syntax level. To address the precedence problem, PEST provides an additional method called "precedence climber" that can rebuild the precedence level after parsing. Readers should be able to find the documentation of "precedence climber" in the PEST cookbook. Until I was writing this paper, the documentation for "precedence climber" was only an empty section with a title. It is possible to read the source code of PEST to find out how to use "precedence climber", though that is relatively hard.

\section{Value and target evaluation}
\label{subsection:value_and_target_evaluation}
This section illustrates the process of evaluating values and language elements. This section focuses on explaining the evaluated code structure, which is important for future Tydi-lang development and debugging errors in Tydi-lang source code. Two optimization methods are applied during the evaluation. The first optimization method is called "lazy evaluation", which means the compiler only evaluates the values that have been used. The second optimization method is called "multi-threading evaluation", indicating that we can use multi-threading to accelerate the evaluation process.

\subsection{Evaluated code structure}
\label{subsubsection:evaluated_code_structure}
The following snippet shows an example of "code structure \#1", the direct result after transforming from AST. The meaning of terms is explained in the comments. Please notice that the comments are manually added rather than a part of the syntax.

\begin{lstlisting}[language=c++]
//corresponds to code structure #1
Project(test_project){          //project name
  Package(tpch){                //package name
    Scope(package_tpch){        //package scope
      Variables{                //variables in this scope
        max_decimal_15:UnknownType(NotInferred("10^15 - 1"))                //because nothing is evaluated, so everything is NotInferred
        day_max:UnknownType(NotInferred("31"))                              //raw integer expression is also not evaluated
        $package$tpch:PackageType(NotInferred(""))
        year_max:UnknownType(NotInferred("10^5 - 1"))
        month_max:UnknownType(NotInferred("12"))
        bit_width_decimal_15:UnknownType(NotInferred("ceil(log2(max_decimal_15))"))
      }
      Types{
        key_stream:VarType(int_stream)          //this represents a type alias
        Date:DataGroup(Date){                   //this is a logical group type
          Scope(group_Date){                        //the scope of the logical group
            Types{
              year:VarType(year_t)                  //the logical types inside the logical group type, the logical type is a reference of identifier "year_t"
              month:VarType(month_t)                
              day:VarType(day_t)
            }
            ScopeRelations{
              --GroupScope-->package_tpch           //scope relationship
            }
          }
        }
        day_t:Bit(NotInferred("ceil(log2(day_max))"))
        date_stream:Stream(date_stream){
          DataType=VarType(Date)
          dimension=NotInferred("1"), user=DataNull, throughput=1, synchronicity=Sync, complexity=7, direction=Forward, keep=false
        }
      }
      Streamlets{
        ...
        Streamlet(region_s)<NormalStreamlet>{
          Scope(streamlet_region_s){
            ScopeRelations{
              --StreamletScope-->package_tpch
            }
            Ports{
              r_comment:Port(VarType(varchar_stream),out) `DefaultClockDomain       //streamlet ports, the port type is a reference of logical type "varchar_stream".
              r_regionkey:Port(VarType(key_stream),in) `DefaultClockDomain
              r_name:Port(VarType(SQL_char25_stream),out) `DefaultClockDomain
            }
          }
        }
        ...
      }
      Implements{
        ...
        Implement(data_filter_i)<NormalImplement> -> ProxyStreamlet(data_filter_s<>){       //proxyStreamlet indicates this should be a streamlet but not evaluated yet.
          Scope(implement_data_filter_i){
            ScopeRelations{
              --ImplementScope-->package_tpch
            }
            Instances{
              l_extendedprice_filter:(NotInferred("stream_filter_1bit_i"))      //not inferred implementation
                ...
            }
            Connections{
              Self.NotInferred("o_shippriority_in") =0=> ExternalOwner(o_shippriority_filter).NotInferred("input") (connection_14449-14498)  //a connection with not inferred ports
                ...
            }
          }
          simulation_process{None}
        }
        ...
      }
    }
  }
}

\end{lstlisting}

After evaluation and template expansion, the corresponding code structure would transform to the following format. Please notice that anything that remains in "NotInferred" state means it is not used in the code. 

\begin{lstlisting}[language=c++]
//corresponds to code structure #4
Project(test_project){          //project name
  Package(tpch){                //package name
    Scope(package_tpch){        //package scope
      Variables{                //variables in this scope
        max_decimal_15:int(999999999999999)             //the value of all variables are calculated and inferred
        day_max:int(31)
        $package$tpch:PackageType(NotInferred(""))      //the package variable will not be evaluated
        year_max:int(99999)
        month_max:int(12)
        bit_width_decimal_15:int(50)
      }
      Types{
        key_stream:Stream(int_stream){
          DataType=Bit(32)
          dimension=1, user=DataNull, throughput=1, synchronicity=Sync, complexity=7, direction=Forward, keep=false
        }           //key_stream is an alias of "int_stream", and the content of the "int_stream" is printed out
        Date:DataGroup(Date){
          Scope(group_Date){
            Types{
              year:Bit(17)      //the bit width of each logical type is evaluated
              month:Bit(4)
              day:Bit(5)
            }
            ScopeRelations{
              --GroupScope-->package_tpch
            }
          }
        }
        day_t:Bit(5)
        date_stream:Stream(date_stream){
          DataType=DataGroup(Date)
          dimension=1, user=DataNull, throughput=1, synchronicity=Sync, complexity=7, direction=Forward, keep=false
        }
      }
      Streamlets{
        ...
        Streamlet(region_s)<NormalStreamlet>{
          Scope(streamlet_region_s){
            ScopeRelations{
              --StreamletScope-->package_tpch
            }
            Ports{
              r_comment:Port(Stream(varchar_stream),out) `DefaultClockDomain        //the logical type is evaluated and the name is the direct name of the logical type
              r_regionkey:Port(Stream(int_stream),in) `DefaultClockDomain
              r_name:Port(Stream(SQL_char25_stream),out) `DefaultClockDomain
            }
          }
        }
        ...
      }
      Implements{
        ...
        Implement(data_filter_i)<NormalImplement> -> Streamlet(data_filter_s){  //the streamlet identifier is evaluated
          Scope(implement_data_filter_i){
            ScopeRelations{
              --ImplementScope-->package_tpch
            }
            Instances{
              l_extendedprice_filter:(Implement(stream_filter_1bit_i@Stream(SQL_decimal_15_2_stream)))  //the identifier "stream_filter_1bit_i@Stream" is a streamlet expaned from a template.
              selection:(Implement(where_claus_i))      //implementation identifier is evaluated
                ...
            }
            Connections{
                ...
              Self.o_shippriority_in:Port(Stream(int_stream),in) `DefaultClockDomain =0=> ExternalOwner(o_shippriority_filter).input:Port(Stream(int_stream),in) `DefaultClockDomain (connection_14449-14498) 
                ...
            }
          }
          simulation_process{None}
        }
        ...
      }
    }
  }
}
\end{lstlisting}

In addition, as mentioned in the Tydi-lang specification, templates will not be evaluated, so in the code structure, a template should seem to be not evaluated. The following code snippet gives an example.

\begin{lstlisting}[language=c++]
//a template will not be evaluated during evaluation
        Streamlet(accumulator_s)<@LogicalDataType(DummyLogicalData)>{       //a template which accepts a logical type as template argument. The "DummyLogicalData" indicates it is a place holder. 
          Scope(streamlet_accumulator_s){
            Variables{
              data_type:DummyLogicalData(NotInferred("$arg$data_type"))     //argument logical type
            }
            Types{
              count_type:Stream(count_type){
                DataType=Bit(NotInferred("32"))
                dimension=0, user=DataNull, throughput=1, synchronicity=Sync, complexity=7, direction=Forward, keep=false
              }
              overflow_type:Stream(overflow_type){
                DataType=Bit(NotInferred("1"))
                dimension=0, user=DataNull, throughput=1, synchronicity=Sync, complexity=7, direction=Forward, keep=false
              }
            }
            ScopeRelations{
              --StreamletScope-->package_tpch
            }
            Ports{
              overflow:Port(VarType(overflow_type),out) `DefaultClockDomain
              count:Port(VarType(count_type),out) `DefaultClockDomain
              output:Port(VarType(data_type),out) `DefaultClockDomain
              input:Port(VarType(data_type),in) `DefaultClockDomain
            }
          }
        }
\end{lstlisting}

The "for" and "if" blocks will be removed from the code structure after expansion. The compiler will perform expansion while evaluating the implementation). Thus the "for"/"if" expansion will be performed after template expansion.

\subsection{Lazy evaluation}
\label{subsubsection:lazy_evaluation}
The term "lazy evaluation" in Tydi-lang means the language elements will not be evaluated if they are not used. This feature applies to all "evaluation" operations in Figure \ref{fig:tydi_lang_frontend_work_flow}, and is designed to avoid the excessive compiling time waste introduced by the standard library and templates. Another benefit of the lazy evaluation is enabling the compiler to dim the unused variables, types, and templates. Whether these language elements are used or not can be determined by looking up the evaluated code structure. 

\begin{figure}[H]
    \centering
     \includegraphics[width=\columnwidth]{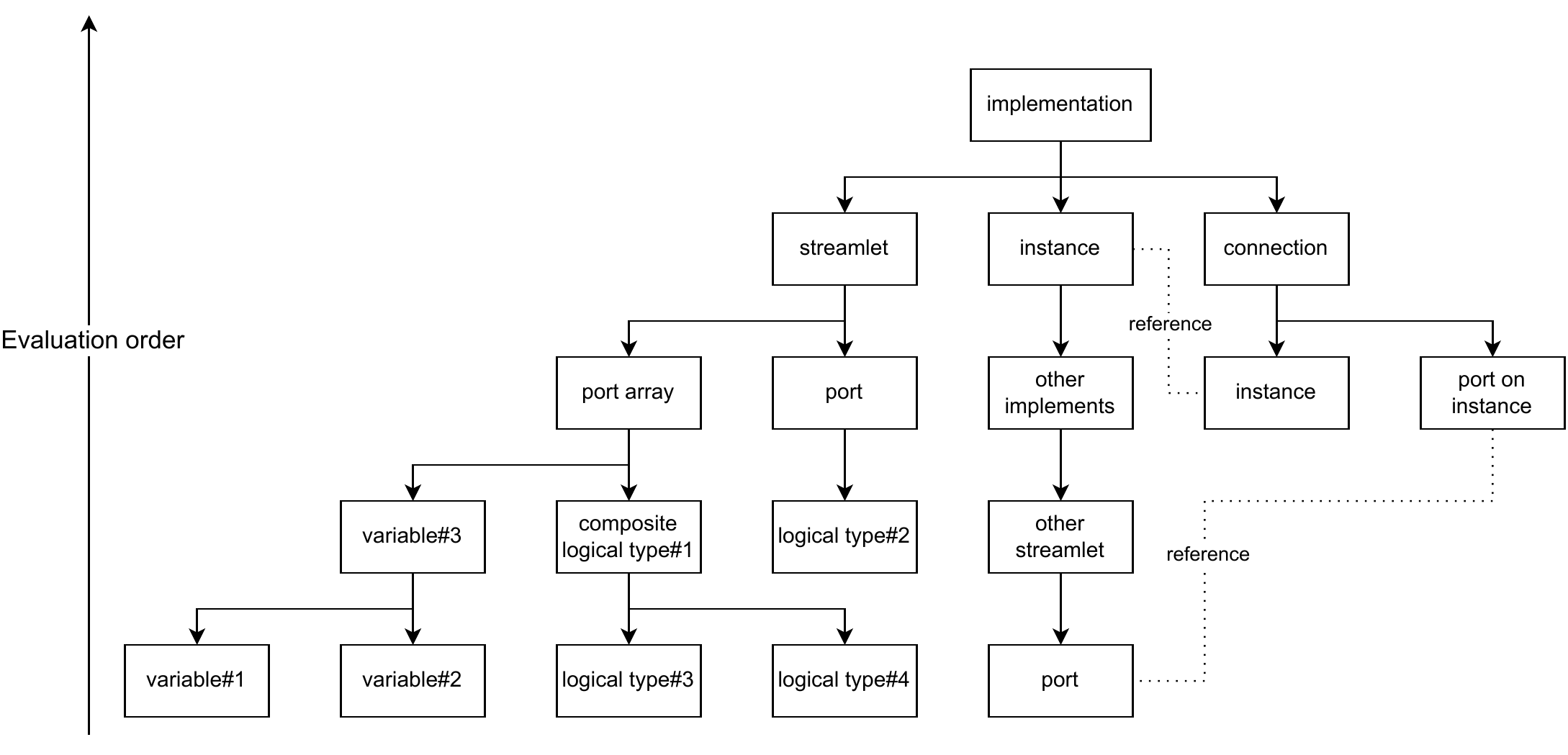}
    \caption{Lazy evaluation of an implementation}
    \label{fig:dependency_tree}
\end{figure}

Figure \ref{fig:dependency_tree} shows the lazy evaluation process, which starts by analyzing the dependency tree of each implementation, excluding implementation templates. The layer-1 leaf nodes of the dependency tree include streamlet, instance, and connection, which are the direct child elements, and the layer-2 leaf nodes include port, other implementations, etc. The bottom layer of the tree consists of variables and logical types, which are the starting places of evaluation. This evaluation pattern ensures that all evaluated elements are used, and unused elements will never be evaluated.

\subsection{Multi-thread evaluation}
\label{subsubsection:multi_thread_evaluation}
Each implementation has its dependency tree. Multiple implementations result in a directed acyclic graph (DAG) whose root nodes are implementations. As mentioned in Section \ref{subsubsection:Mutable_memory_structure_in_Rust}, all nodes in the dependency tree are protected by locks. Evaluating the DAG can be optimized with multi-threading because each implementation is an entry for a thread, and there are multiple implementations. However, several threads might repeat evaluating the same nodes because they are common leaf nodes. Thus, locks are applied to avoid repeat evaluation. Deciding whether a lock should protect this element depends on how many leaf nodes exist under it. Using locks on high-level elements, such as streamlets, will result in coarse-grained parallelism: Vice versa, fine-grained parallelism requires locking low-level elements such as variables. The current Tydi-lang compiler chooses a compromise, which uses locks on child nodes of streamlets.

Currently, the multi-threading evaluation is only implemented but not tested yet. The reason is that multi-file parsing is impossible because the backend does not support multi-file generation. The performance improvement of multi-thread evaluation is not measured because the project structure is not complicated enough, causing the implementation dependency structure to be more like a linear structure.

\section{Tydi standard library}
\label{subsubsection:Tydi_standard_library}

The Tydi-lang standard library is a pure-template library, defining many frequently-used components which can be categorized into the following three types.

\begin{itemize}
    \item Components to duplicate/remove data packets. The Tydi-lang is designed for streaming hardware where each port can only be connected once, while using a value several times is common in software languages. Thus duplicator and voider (a component name) are proposed to duplicate data packets and remove data packets. In the low-level implementation, duplicators copy and resend the bit-level data to all output ports and only acknowledge the input port when all outputs are acknowledged. Voiders will remove all data packets by acknowledging the source component and ignoring the data. These two components work on the handshaking layer and hardware bit, so they are templates in Tydi-lang. 
    \item Components that describe common behaviors for different logical types. For example, an adder can work for integer types, decimal types, and many other numerical types once the bit width is specified. A comparator is also possible to compare integers, dates, etc. However, selecting and implementing these components might be tricky because the multipliers for integer and decimal are different (if taking the digits after the digit point into consideration). For this case, assertion and "if" can be applied to restrict the template.
    \item Components to transform logical types. The transformation includes splitting a group type into its inner types or combining several logical types in a group. These template components help process ports with user-defined composite data structures. This part is future work and has not been implemented in the current Tydi-lang version.
\end{itemize}

Unlike typical template components, the components in the Tydi-lang standard library are too elementary to be described as instances and connections (external implementations if using terms in Table \ref{table:tydi_ir_terms} ), so there is another RTL generation process for these standard components. However, this generation process must be manually defined. For example, in a duplicator template with two arguments - a logical type of stream and an integer variable to indicate the output port count, the process to generate the correct component needs to be hardcoded into the generator.

Because adding a new component template in the Tydi standard library means adding more hard-coded processes in the generator, the Tydi-lang standard library should be kept as small and as abstract as possible. It is a compromise between the library size and the generator complexity, resulting in greater difficulty in designing the standard library. In addition, finding the proper abstraction of each component is also complicated. The selection of components in the Tydi-lang library and their corresponding templates remain under construction. The Tydi-lang library used in Chapter \ref{chapter:result_and_evaluation} is a prototype and only includes the essential templates for our test cases. 

\section{Sugaring}
\label{subsubsection:sugaring}

Sugaring is important in reducing language developers' design effort by automatically inferring and appending the absent code. With the help of the Tydi-lang standard library, the current compiler provides two types of sugaring. The first type of sugaring is the automatic duplicator template insertion if an output port has been connected to multiple input ports. The compiler will automatically infer the duplicator template's logical type and output channel size. The second type of sugaring is the automatic voider template insertion if an output port has never been used, where voider is a component that does nothing but is always ready to receive the next packet. These two sugarings release the restriction that "one port must be connected to exactly one other port".

For sugaring examples, consider the case of using Fletcher \cite{fletcher} to generate components to access memory data from a data schema. The data schema might be large while the query only accesses a small portion of it, and the query on data is flexible while the generated components are rigid. Without sugaring, developers need to manually append voiders for each unused port on the generated Fletcher components. Another example can be found in the translation from software programming languages to Tydi-lang. In software programming languages, using variables multiple times is normal because it is a value inside the memory that can be accessed at any time. However, for hardware design, the value is represented by logical gate states, which are transient. Hardware designers usually manually duplicate the stream to send data to multiple components. Sugaring in Tydi-lang can automatically put duplicators between the source port and sink port according to the times that the data is used in the Tydi-lang source code. 

\section{Design rule check}
\label{subsubsection:design_rule_check}
Design rule checks (DRC) are widely applied in hardware designing areas, from PCB design to IC design. Tydi-lang is special due to its high-level properties, so a specialized Tydi-specific DRC system is integrated with the compiler, aiming to find out high-level design errors. Low-level DRC is still necessary and can be performed after generating low-level HDLs. Tydi-IR also integrates a DRC system to check design errors. The difference among the Tydi-lang DRC and Tydi-IR DRC is that Tydi-IR always checks the logical types of two ports of a connection have the same type hierarchy, while Tydi-lang provides two options to compare the type equality.

The design rule check is performed in the last stage of the compiling process to find out high-level errors which become invisible after generating low-level HDLs. It checks the following rules:

\begin{itemize}
    \item The logical types of the connected ports are compatible. Compatible means the logical types refer to the same logical type or the two logical types have the group/union structure and same bit-width of their children. The syntax to select one of the two compatibility rules is mentioned in Section \ref{subsubsection:Implement}.
    \item A connection is established from a source(output) port to a sink(input) port. Please notice that the direction of an input port is output for that implementation and input for other implementations.
\end{itemize}

More rules can be added in the future, e.g. checking the port complexities are compatible. The Rust source files to define these checks is available in the following link:

\href{https://github.com/twoentartian/tydi-lang/blob/main/tydi_lang_front_end/src/drc.rs}{tydi-lang/tydi\_lang\_front\_end/src/drc.rs}

%% file: 05-tydi-simulator.tex
\chapter{Tydi simulator}

\section{Introduction to Tydi simulator}


The goal of Tydi-lang simulator is assisting high-level developers in designing streaming circuit to meet functional requirements regardless of low-level behavior, and generating testbenches to collaborate with low-level developers.

Simulating the streaming hardware on the Tydi-lang level is necessary because the response time of a single component is determined by the arrival time of asynchronous input data packets. Analyzing the timing information of all components can quickly help designers identify streaming bottlenecks. Using traditional low-level simulators for such work is cumbersome because there are too many trivial low-level signals such as handshaking. Our simulator can also predict the output sequences under certain input sequences, but this is also possible with traditional simulation tools, so we will not address this feature in this thesis.

Performing simulation requires the input data sequence to top-level implementation and the mapping from the clockdomain to physical frequency and phase. The simulator can calculate the delay time, record data flows, and record the state-transition table of each implementation.

The delay time includes the delay from components simulation code and connection. The time to transfer data packets via connections is calculated with the connection clockdomain and data packet length. The data flow and the state transformation can be inferred from the simulation code. The state means the combination of all possible values of all state variables. Notice that some hardware components cannot be described by the "state" system, for example, the random number generator. 

Because state transformation is caused by events, which are combinations of receiving data from different ports, analyzing the relationship between data flow and state could also help identify the potential for deadlock. As for identifying bottlenecks, the simulator should be able to record the waiting time of all output ports (blocked by handshaking). Designers can investigate the output ports with the longest blockage to find the bottleneck component.

\section{Tydi simulation syntax}
\label{subsection:tydi_simulation_syntax}
The Tydi-lang simulation code is defined inside an implementation to describe its behavior. Implementation defined by inner instances and connections should not have simulation code because inner instances characterize its behavior. The simulation syntax includes the following parts.

\begin{itemize}
    \item State variable: represents a state with a string value.
    \item Acknowledge mechanism: because the Tydi-lang integrates the handshaking mechanism from Tydi-spec, it is crucial to control the handshaking behavior and time. For example, a component with two input ports with different throughputs should have synchronization on its ports. This synchronization can be achieved by controlling the time of acknowledging the output ports.
    \item Event-driven: an event is an action from ports, such as receiving a data packet. Designers can use boolean logic to define composite events. For example, only compute when both data from two ports are ready. The process when an event happens is called an event handler, where behavior code, such as sending acknowledge signals, changing state variables, sending data to other components, and delaying for a specific time, can be defined here. In addition, the "if" and "for" syntax is available in the event handler as logic flow control syntax.
\end{itemize}

The simulation syntax is not finished yet. Currently, the simulation code can be parsed to the correct abstract syntax tree (AST). Readers can find the complete grammar in the aforementioned PEST file. Here I provide a proposed simulation example below.


\begin{lstlisting}[language=c++]
impl impl_template<i:int> of basic0 {
  instance test_inst(basic0_1) [i],

  process {
    state component_state = "0"; //declare state variable "component_state" and its initial state as "0"
    set_ack(data_in_0, 2);      //set the acknowledge count of port "data_in_0" to 2
    set_ack(data_in_1, 1);

    event receive(data_in_0) && receive(data_in_1) {
      if (component_state == "0") {
        delay_cycle(5, 100MHz);
        send(data_out_0, 0b11110000);
        //do we need read(data_in_0)?
        read(data_in_0);
        //for composite data types: Group(a: Bit(8), b: Bit(8))
        send(data_out_0, Group(a=0x11110000, b=0x11110000));
        send(data_out_0, Union(a=0x11110000));
        //for composite data types: Group(a: Bit(8), b: Stream(Bit(8)))
        send(data_out_0->b, 0x11110000);

        assign component_state = "1";   //assign the state variable "component_state" to state "1"
      }
      elif (component_state == "1") {
        assign component_state = "2";
      }
      elif (component_state == "2") {
        assign component_state = "1";
      }
      ack(data_in_1);
      ack(data_in_0);
    };

    event receive(data_in_0) {
      ack(data_in_0);
    };
  },
};
\end{lstlisting}

The simulation block is started by the keyword "process". In the simulation block, users can define state variables, set the acknowledge count and declare events. The acknowledge count is a mechanism to determine when to acknowledge the source port. For the above example, the acknowledge count of "data\_in\_0" is set to 2, each "ack" statement in lines 30 and 34 will add 1 to the counter. The source port will be acknowledged when the counter reaches 2. Each event is a logical expression of one or multiple built-in functions. For the above example, "receive" is such a built-in function. The event is a new block where users can define the simulated behavior with if/for/built-in functions. Please notice that though there are "if" and "for" syntax in simulation code, their compiling processes are entirely different from generative "if" and "for" syntax because they work on variable values rather than generating parallel components. Due to this reason, the simulator should integrate a small stack-based virtual machine to execute the simulation code. The stack implementation should include a PC(program counter to record the execution location) and a SP (stack pointer, necessary in nested "if"/"for" structure). LR(link register) is not necessary because the event does not return any value. Each event in implementation might bind to multiple stacks because a single event might be triggered multiple times in real hardware. This property also results in the state variables being shared among different stacks, but local variables should be stack-independent. 

One thing that has not been designed in Tydi-lang simulation syntax is the composite data representation, which will be used in "send", "read" functions, and comparison-related features. Using "Group" and "Union" to describe the data structure is also not an efficient solution for developers and should be re-designed.

The simulation syntax also integrates many built-in functions such as \texttt{send}, \texttt{receive}, \texttt{ack}. The decisions about their arguments and return values are not determined, either. Appendix \ref{table:proposed_builtin_functions_tydi_simulator} shows a draft of current design and can be used for reference.

\section{Tydi simulator structure}
\begin{figure}[htbp]
    \centering
     \includegraphics[width=\textwidth]{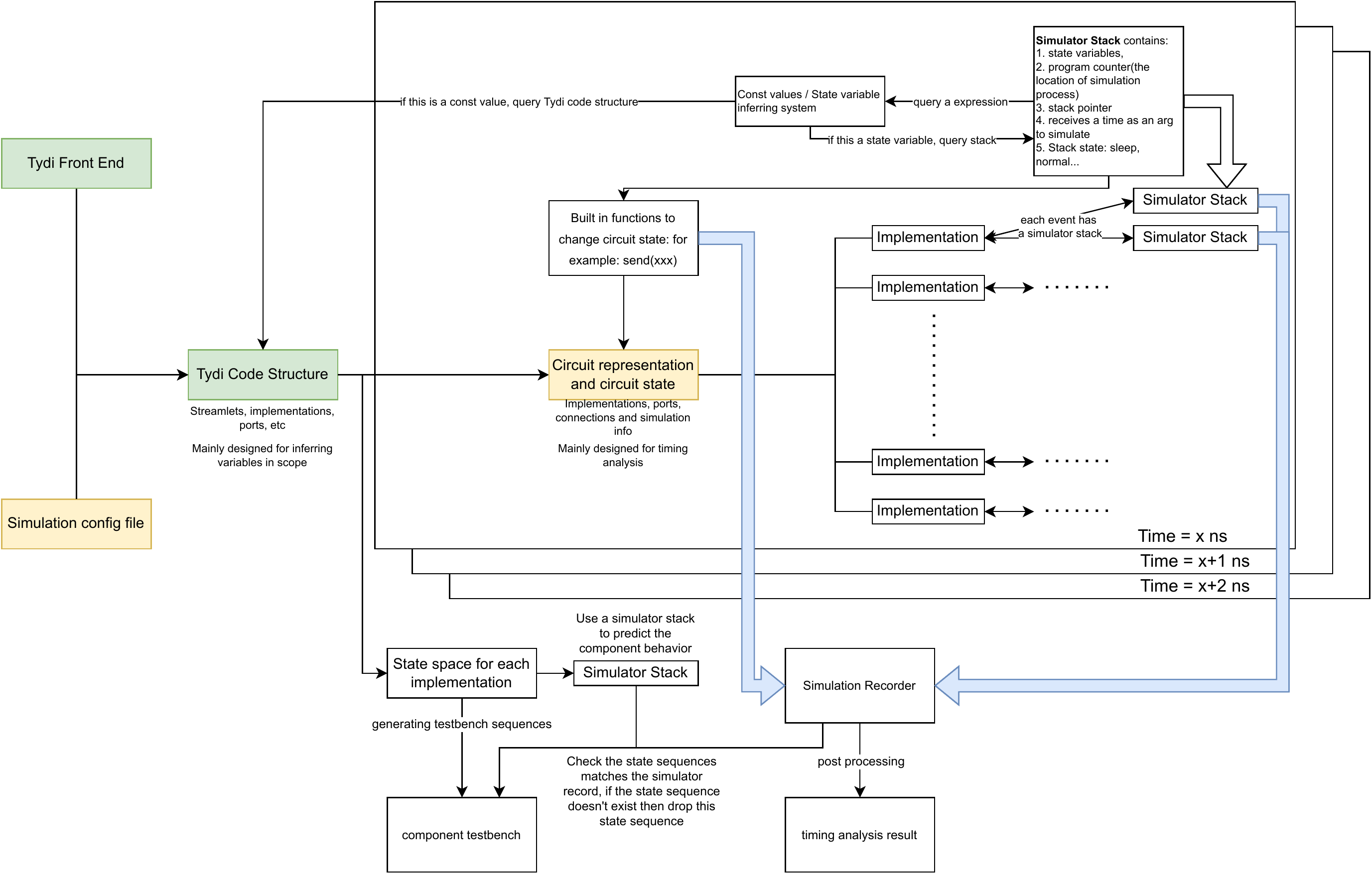}
    \caption{The block diagram of the Tydi-lang simulator}
    \label{fig:tydi_simulator}
\end{figure}

Figure \ref{fig:tydi_simulator} shows the structure of the Tydi simulator. The green parts are already finished, and the yellow parts are partially finished. The white parts indicate that they are not started yet. 

The "Tydi code structure" is a tree-like structure unsuitable for simulation circuits. So it needs to be transformed to a flat circuit representation first. Meanwhile, the simulator should attach the state variables to each implementation. The simulation is performed by a stack-based virtual machine as mentioned in \ref{subsection:tydi_simulation_syntax}. Above all, for a given input data sequence specified in the "simulation config file", the Tydi-lang should be able to provide the predicted output data sequence.

During simulation, a "simulation recorder" will record the state transformations and their corresponding triggered events. In simulation syntax, users will define many transformations, but not all of them will appear in the simulation. Especially with given inputs, the state sequence usually only has limited patterns. The simulation recorder is applied to find out what state patterns are within the design scope. The "state space for each implementation" is directly analyzed from the Tydi-lang simulation code, containing some state patterns outside the design scope. The final testbenches are generated based on the mixture of the simulation recorder and state space. In addition, the timing result can be calculated based on the data sequence stored in the simulation recorder.

\section{Generate testbench}
While the simulation code only describes the expected behavior of components, it does not guarantee low-level behavioral correctness. The Tydi-lang simulator should be able to generate testbench files to ensure the expected behavior matches the low-level simulation results. Tydi-IR already defined a testbench syntax based on prediction strategy (giving certain input and verifying output correctness), and provided a tool to translate from Tydi-IR testbenches to VHDL testbenches. The Tydi-lang simulator can utilize this tool to generate VHDL testbenches.

The mechanism to generate testbench can be briefly described as an "input - current state - output" testing system. The "input" corresponds to an event in Tydi-lang, the "current state" is a combination of events and the initial state, and the "output" corresponds to sending data. Generating testbenches is a process of using the above mechanism to cover all states and events in the state transition table. The coverage of input data in the simulation stage is important because uncovered input results in uncovered state transformation. The testbench system also reduces design effort because only low-level components require simulation code and testbenches, which is easier than writing testbenches for high-level components.

The Tydi-lang testbench system also allows the collaboration between Tydi-lang developers and low-level HDL developers. Tydi-lang developers can focus on using the Tydi toolchain to design streaming applications, whose requirements always change in software domains, while low-level HDL developers can focus on designing and optimizing low-level components, regardless of high-level function requirements. Tydi-lang designers can update simulation code when low-level components are optimized. Low-level language developers can use the testbench from the Tydi-lang toolchain to ensure the correct component behavior.

\section{Current simulator implementation and circuit representation}

This section elaborates on the process of converting Tydi-lang to circuit representation, the only work I have done for the Tydi-lang simulator. Some other drafts about the configuration format are also mentioned here. The Rust source code for this section is in the following link: 

\href{https://github.com/twoentartian/tydi-lang/tree/main/tydi_simulator}{tydi-lang/tydi\_simulator}

The configuration file for Tydi-lang simulator is in JSON format. The configuration file defines the top-level implementation and the input signals on the top-level implementation. For timing analysis, the real frequency and phase of each clockdomain must also be specified.

I rewrite the memory structure for all language elements because we previously focused on evaluating their values and reference relationship, which are useless information in circuit representation. Meanwhile, the memory structure for the simulator requires new features, for example, the references to the two ports for a connection. These new memory structures are written in files starting with "circuit" in the simulator folder. For example, \href{https://github.com/twoentartian/tydi-lang/blob/main/tydi_simulator/src/circuit_connection.rs}{tydi-lang/tydi\_simulator/src/circuit\_connection.rs} defines the memory structure for a connection.

After converting the Tydi-lang code structure to the flat circuit structure, the simulator will generate a DOT language\cite{dot-lang} file. DOT language is a graph description language that uses a special syntax to describe various graphs. The DOT language file is available in the output folder with other compiler outputs. DOT language files can be converted to vector images by extension in Visual Studio Code \cite{dot-lang-vscode-ext}. Chapter \ref{chapter:result_and_evaluation} provides a sample circuit image. 

The unit test "test\_process\_sample\_code" in \href{https://github.com/twoentartian/tydi-lang/blob/main/tydi_simulator/src/test.rs}{tydi-lang/tydi\_simulator/src/test.rs} is an example to automatically generate the DOT language file.

The following code snippet shows some basic syntax rules for DOT language. Notice that the comments are not allowed in DOT language, and these C-style comments are just for explaining. It will not work if you directly copy the code to DOT files.

\begin{lstlisting}[language=c++]
digraph {                   //represents this is a directed graph
main_i [color=red, shape=record, label="{<component>main_i|<err>err|<l_linenumber>l_linenumber|<l_orderkey>l_orderkey|<p_partkey>p_partkey|<revenue>revenue}"];        //top level component is main_i, it has serveral ports: err, l_linenumber, etc. The name in <...> is the reference that will be used to create connection
main_i__accu [shape=record, label="{<component>main_i__accu|<count>count|<input>input|<output>output|<overflow>overflow}"]; //two consecutive "_" means the original component hierarchy.
main_i__data_filter__selection__duplicate_l_shipinstruct_0_output [shape=record, label="{<component>main_i__data_filter__selection__duplicate_l_shipinstruct_0_output|<input>input|<output_AT_0>output@0|<output_AT_1>output@1|<output_AT_2>output@2}"];       //the hierarchy for this component: main_i -> data_filter -> selection -> duplicate_l_shipinstruct_0_output. Because in circuit representation everything is flat, we encode this hierarchy in names for potential future use.

...

main_i__err_and:output -> main_i:err [label="connection_25562-25599__main_i::err_and__main_i"] ;        //Make a connection from the port (output) of component (main_i->err_and) to port (err) of component(main_i)
}

\end{lstlisting}

%% file: 06-evaluation-result.tex
\chapter{Result and evaluation}
\label{chapter:result_and_evaluation}


This chapter provides a use case of applying FPGAs to accelerate SQL queries to demonstrate the increased hardware abstraction level and the decrease in design effort. We translated several TPC-H \cite{tpch} SQL benchmark queries to Tydi-lang to represent the query logic on hardware and compare the line of code (LoC) of Tydi-lang and the generated VHDL. 

As mentioned previously, 
the Tydi-lang integrates a standard library. The code of the standard library should not be counted in to design effort because they can be reused. In big data analytic area, there are tools (such as Fletcher \cite{fletcher}) to automatically generate VHDL hardware interfaces to access memory data. Because there currently are no tools to automatically generate a Tydi-lang interface, the code to describe interfaces is manually written. The primary key in a TPC-H dataframe will be treated as the input port, and the other ports will be treated as output ports. This part of code should not be counted in to design effort either because they can be automatically generated. So in total there are three parts of code in our TPC-H examples: the Tydi-lang standard library, the interface part and query logic part. The line of code (LoC) of each part is counted as a representation of design effort.

\begin{figure}
    \centering
     \includegraphics[width=\textwidth]{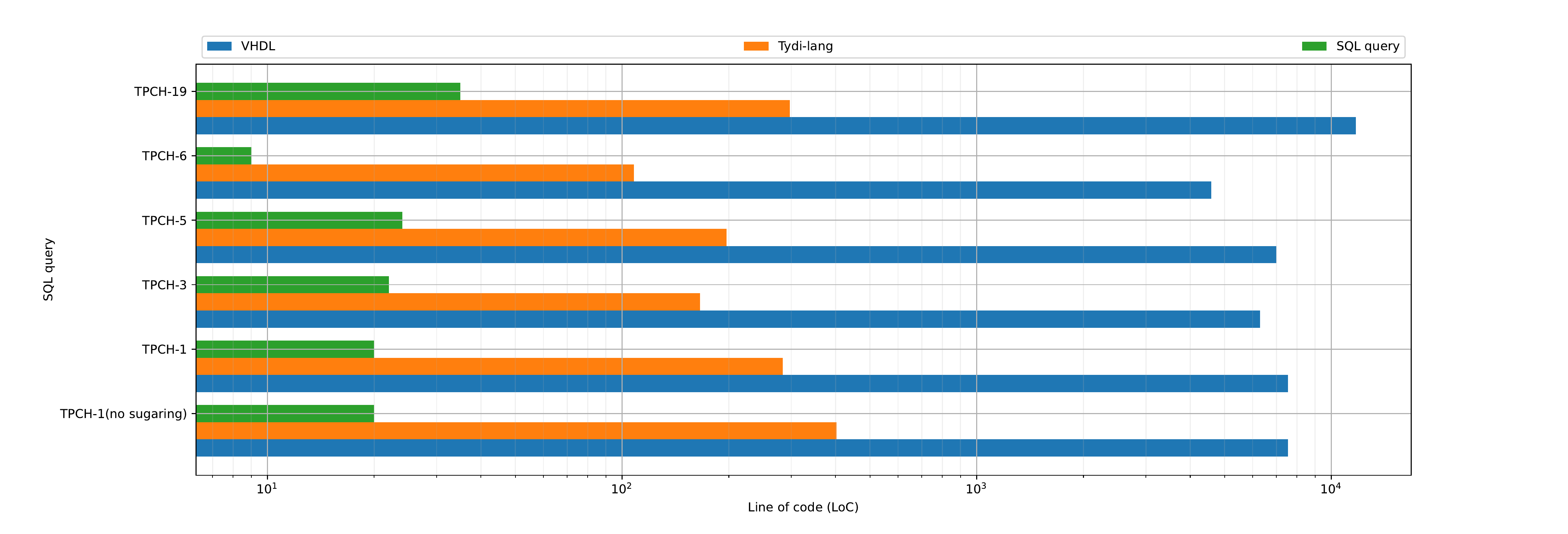}
    \caption{LoC for translating TPC-H queries to Tydi-lang}
    \label{fig:tpch_loc}
\end{figure}

In addition, the result also contains a non-sugaring version of the first query in TPC-H to show the design effort saved by sugaring. The result is shown in Table \ref{table:result} and Figure \ref{fig:tpch_loc}. The following formula presents the calculation of ratio and total LoC.

\[ LoC_a = LoC_q + LoC_f + LoC_s \]
\[ R_q = LoC_{vhdl} / LoC_q \]
\[ R_a = LoC_{vhdl} / LoC_a \]

The Tydi-lang source code, SQL source code, evaluation result, Tydi-IR, and generated VHDL are available in the following link: \href{https://github.com/twoentartian/tydi-lang/tree/main/CookBook}{https://github.com/twoentartian/tydi-lang/tree/main/CookBook}.

\begin{table}[H]
\centering
\caption{LoC for translating TPC-H queries to Tydi-lang}
\label{table:result}
\resizebox{\textwidth}{!}{
\begin{tabular}{|lc|c|c|cc|c|}
\hline
\multicolumn{2}{|r|}{LoC for Fletcher part($LoC_f$)} & \multicolumn{1}{l|}{166} & \multicolumn{1}{l|}{} & \multicolumn{2}{r|}{LoC for Tydi-lang standard library($LoC_s$)} & \multicolumn{1}{l|}{151} \\ \hline \hline

\multicolumn{1}{|c|}{Query name} & Raw SQL query & \begin{tabular}[c]{@{}c@{}}Query logic \\ in Tydi-lang \\ ($LoC_q$) \end{tabular} & \begin{tabular}[c]{@{}c@{}}Total Tydi-lang LoC \\ ($LoC_a$) \end{tabular} & \multicolumn{1}{c|}{\begin{tabular}[c]{@{}c@{}}Generated VHDL\\ ($LoC_{vhdl}$)\end{tabular}} & \begin{tabular}[c]{@{}c@{}}Ratio: \\ VHDL/Query logic\\ ($R_q$)\end{tabular} & \begin{tabular}[c]{@{}c@{}}Ratio:\\ VHDL/Total Tydi-lang\\ ($R_a$)\end{tabular} \\ \hline
\multicolumn{1}{|l|}{TPC-H 1 (without sugaring)} & 20 & 402 & 709 & \multicolumn{1}{c|}{7547} & 18.77 & 10.50 \\ \hline
\multicolumn{1}{|l|}{TPC-H 1} & 20 & 284 & 601 & \multicolumn{1}{c|}{7547} & 26.57 & 12.56 \\ \hline
\multicolumn{1}{|l|}{TPC-H 3} & 22 & 166 & 483 & \multicolumn{1}{c|}{6291} & 37.90 & 13.02 \\ \hline
\multicolumn{1}{|l|}{TPC-H 5} & 24 & 197 & 514 & \multicolumn{1}{c|}{6992} & 35.49 & 13.60 \\ \hline
\multicolumn{1}{|l|}{TPC-H 6} & 9 & 108 & 425 & \multicolumn{1}{c|}{4586} & 42.46 & 10.79 \\ \hline
\multicolumn{1}{|l|}{TPC-H 19} & 35 & 297 & 614 & \multicolumn{1}{c|}{11734} & 39.51 & 19.11 \\ \hline
\end{tabular}%
}
\end{table}

Other queries in TPC-H benchmark are not translated into Tydi-lang because some of them have nested "select" structures, which requires storing the intermediate result back to memory for later calculations. The interface of storing and accessing intermediate result is beyond the research scope of Tydi-lang.

The result shows that using Tydi-lang can greatly reduce the number of lines of code to design FPGA accelerators. If we use the $R_q$ as the indicator of design effort (the code in the standard library and memory interface does not count), the total design effort can be saved for over 40x in TPC-H 6 query. 
The reduction in LoC comes from many aspects. For the Tydi-lang frontend, the reduction mainly comes from the following points.

\begin{itemize}
    \item The Fletcher and Tydi-lang standard library provide many component templates.
    \item Using templates can get higher $R_q$ because many components, such as comparators and constant data generators, are generated from templates.
    \item Desugaring process can automatically add missing components, such as voiders and stream duplicators.
\end{itemize}

The Tydi-lang backend can also reduce the line of code because the Tydi type system can encode many ports in a single Tydi type and many connections in a single Tydi connection.

Different queries can get different $R_q$ due to some intrinsic properties. The most interesting example is the TPC-H query 19, which can get relatively higher $R_q$ because it contains similar sub-structures. The SQL source code of TPC-H query 19 is provided below.

\begin{lstlisting}[language=sql]
:x
:o
select
	sum(l_extendedprice* (1 - l_discount)) as revenue
from
	lineitem,
	part
where
	(
		p_partkey = l_partkey
		and p_brand = ':1'
		and p_container in ('SM CASE', 'SM BOX', 'SM PACK', 'SM PKG')
		and l_quantity >= :4 and l_quantity <= :4 + 10
		and p_size between 1 and 5
		and l_shipmode in ('AIR', 'AIR REG')
		and l_shipinstruct = 'DELIVER IN PERSON'
	)
	or
	(
		p_partkey = l_partkey
		and p_brand = ':2'
		and p_container in ('MED BAG', 'MED BOX', 'MED PKG', 'MED PACK')
		and l_quantity >= :5 and l_quantity <= :5 + 10
		and p_size between 1 and 10
		and l_shipmode in ('AIR', 'AIR REG')
		and l_shipinstruct = 'DELIVER IN PERSON'
	)
	or
	(
		p_partkey = l_partkey
		and p_brand = ':3'
		and p_container in ('LG CASE', 'LG BOX', 'LG PACK', 'LG PKG')
		and l_quantity >= :6 and l_quantity <= :6 + 10
		and p_size between 1 and 15
		and l_shipmode in ('AIR', 'AIR REG')
		and l_shipinstruct = 'DELIVER IN PERSON'
	);
:n -1
\end{lstlisting}

We can observe that the three "or" components have the same structure. The only difference is that the argument strings are different. In this case, the three "or" components can be written as a template which receives these string arguments and use "for" statement in Tydi-lang to generate three components (the current $R_q$ for query 19 is lower than expected due to an issue mentioned in Appendix \ref{subsubsection:Duplicated_identifier_issue_in_for_if_expansion}.). The string arguments can be stored in a Tydi-lang array. These abstractions can further save design effort.

I also found that there are fixed patterns to translate SQL to Tydi-lang. For example, the "select" keyword always maps to a "stream\_filter" in Tydi-lang, which receives a data packet and a one-bit signal to determine whether to send this packet to next stage. These patterns can be easily found in the Tydi-lang source code. It might be possible to design an automation tool to translate SQL to Tydi-lang.

Please also notice that the generated VHDL only includes hardware structure because the RTL generator for Tydi-lang standard library is not finished yet (mentioned in Section \ref{subsubsection:Tydi_standard_library}). In the future version with a finished RTL generator, the real $R_q$ would be higher than current result.

Besides the overall LoC comparison, the Tydi-lang source code of TPC-H query 1 is also provided to illustrate a typical Tydi-lang application. The source code does not include the Fletcher part because it should be automatically generated in the future. As for the Tydi-lang standard library part, only the interface (streamlet) code is provided because the implementation should be generated by the code generator of the standard library.

\begin{lstlisting}[language=c++]
package std;

type SQL_int = Bit(32);
type int_stream = Stream(SQL_int, d = 1);
type key_stream = int_stream;

const year_max = 10^5 - 1;
type year_t = Bit(ceil(log2(year_max)));
type year_stream = Stream(year_t);
const month_max = 12;
type month_t = Bit(ceil(log2(month_max)));
type month_stream = Stream(month_t);
const day_max = 31;
type day_t = Bit(ceil(log2(day_max)));
type day_stream = Stream(day_t);
type Group Date {
  year: year_t,
  month: month_t,
  day: day_t,
};
type date_stream = Stream(Date, d = 1);

type SQL_char = Bit(8);
type SQL_char1_stream = Stream(SQL_char, d = 1);
type varchar_stream = Stream(SQL_char, d = 2);

type SQL_char10 = Bit(8*10);
type SQL_char10_stream = Stream(SQL_char10, d = 1);
type SQL_char15 = Bit(8*15);
type SQL_char15_stream = Stream(SQL_char15, d = 1);
type SQL_char25 = Bit(8*25);
type SQL_char25_stream = Stream(SQL_char25, d = 1);

const max_decimal_15 = 10^15 - 1;
const bit_width_decimal_15 = ceil(log2(max_decimal_15));
type SQL_decimal_15 = Bit(bit_width_decimal_15);
type Group SQL_decimal_15_2 {
  const frac = 2,
  decimal: SQL_decimal_15,
};
type SQL_decimal_15_2_stream = Stream(SQL_decimal_15_2, d = 1);

////////////////////  Fletcher part  ////////////////////
...

////////////////////  tpch.part  ////////////////////
...

////////////////////  tpch.nation  ////////////////////
...

////////////////////  tpch.region  ////////////////////
...

////////////////////  tpch.supplier  ////////////////////
...

////////////////////  tpch.partsupp  ////////////////////
...

////////////////////  tpch.customer  ////////////////////
...

////////////////////  tpch.orders  ////////////////////
...

////////////////////  tpch.lineitem  ////////////////////
...

////////////////////  tydi standard lib  ////////////////////

//void component, always acknowledge the handshake
streamlet void_s<type_in: type> {
  input: type_in in,
};

...

//padding zero to the highest bit
streamlet padding_zero_s<type_in: type, type_out: type> {
  stream_in: type_in in,
  stream_out: type_out out,
};

...

//comparator, compare two values: (input0 is larger) => 1, (input1 is larger) => 2, (input1 == input0) => 3
streamlet comparator_s<type_in: type> {
  input0: type_in in,
  input1: type_in in,
  output: Stream(Bit(2)) out,
};

...

//const value generator, type_out should be a Stream(Bit(x)) type and the value should be the value mapped to Bit(x)
streamlet const_value_generator_s<type_out: type, value: int> {
  output: type_out out,
};

...

//data duplicator
streamlet duplicator_s<data_type: type, output_channel: int> {
  input: data_type in,
  output: data_type [output_channel] out,
};

...

//stream filter
type stream_filter_select_stream = Stream(Bit(2));
streamlet stream_filter_s<data_type: type> {
  input: data_type in,
  output: data_type out,
  select: stream_filter_select_stream in,
};

...

//accumulator
streamlet accumulator_s<data_type: type> {
  type count_type = Stream(Bit(32)),
  input: data_type in,
  output: data_type out,
  count: count_type out,
  type overflow_type = Stream(Bit(1)),
  overflow: overflow_type out,
};

...

//logical type converter
streamlet converter_s<input_type: type, output_type: type, channel: int> {
  input: input_type [channel] in,
  output: output_type [channel] out,
};

...

//and
streamlet and_s<data_type: type, input_channel: int> {
  input: data_type [input_channel] in,
  output: data_type out,
};

...

//adder
streamlet adder_s<data_type: type> {
  input0: data_type in,
  input1: data_type in,
  output: data_type out,
  overflow: Stream(Bit(1)) out,
};

...

//to negative
streamlet to_neg_s<data_type: type> {
  input: data_type in,
  output: data_type out,
};

...

//multiplier
streamlet multiplier_s<data_type: type> {
  input0: data_type in,
  input1: data_type in,
  output: data_type out,
  overflow: Stream(Bit(1)) out,
};

...

//divider
streamlet divider_s<data_type: type> {
  dividend: data_type in,
  divisor: data_type in,
  quotient: data_type out,
};

...

////////////////////  Project file  ////////////////////
//construct the sql_date stream by providing its year steam, month steam, and day stream
streamlet sql_date_constructor_s {
  year_input: year_stream in,
  month_input: month_stream in,
  day_input: day_stream in,
  date_output: date_stream out,
};

external impl sql_date_constructor_i of sql_date_constructor_s {

};

streamlet const_date_generator_s {
  date_output: date_stream out,
};

impl const_date_generator_i<day: int, month: int, year:int> of const_date_generator_s {
  instance day_gen(const_value_generator_i<type day_stream, day>),
  instance month_gen(const_value_generator_i<type month_stream, month>),
  instance year_gen(const_value_generator_i<type year_stream, year>),
  instance compositor(sql_date_constructor_i),

  day_gen.output => compositor.day_input,
  month_gen.output => compositor.month_input,
  year_gen.output => compositor.year_input,
  compositor.date_output => date_output,
};

streamlet data_filter_s {
  l_partkey_in: key_stream in,
  l_suppkey_in: key_stream in,
  l_quantity_in: SQL_decimal_15_2_stream in,
  l_extendedprice_in: SQL_decimal_15_2_stream in,
  l_discount_in: SQL_decimal_15_2_stream in,
  l_tax_in: SQL_decimal_15_2_stream in,
  l_returnflag_in: SQL_char1_stream in,
  l_linestatus_in: SQL_char1_stream in,
  l_shipdate_in: date_stream in,
  l_commitdate_in: date_stream in,
  l_receiptdate_in: date_stream in,
  l_shipinstruct_in: SQL_char25_stream in,
  l_shipmode_in: SQL_char10_stream in,
  l_comment_in: varchar_stream in,

  l_partkey_out: key_stream out,
  l_suppkey_out: key_stream out,
  l_quantity_out: SQL_decimal_15_2_stream out,
  l_extendedprice_out: SQL_decimal_15_2_stream out,
  l_discount_out: SQL_decimal_15_2_stream out,
  l_tax_out: SQL_decimal_15_2_stream out,
  l_returnflag_out: SQL_char1_stream out,
  l_linestatus_out: SQL_char1_stream out,
  l_shipdate_out: date_stream out,
  l_commitdate_out: date_stream out,
  l_receiptdate_out: date_stream out,
  l_shipinstruct_out: SQL_char25_stream out,
  l_shipmode_out: SQL_char10_stream out,
  l_comment_out: varchar_stream out,

};

impl data_filter_i of data_filter_s {
  instance baseline_date(const_date_generator_i<1,12,1998>),
  instance compare_date(comparator_i<type date_stream>),
  l_shipdate_in => compare_date.input0,
  baseline_date.date_output => compare_date.input1,

  instance l_partkey_bypass(stream_filter_i<type key_stream>),
  l_partkey_in => l_partkey_bypass.input,
  compare_date.output => l_partkey_bypass.select,
  l_partkey_bypass.output => l_partkey_out,

  instance l_suppkey_bypass(stream_filter_i<type key_stream>),
  l_suppkey_in => l_suppkey_bypass.input,
  compare_date.output => l_suppkey_bypass.select,
  l_suppkey_bypass.output => l_suppkey_out,

  instance l_quantity_bypass(stream_filter_i<type SQL_decimal_15_2_stream>),
  l_quantity_in => l_quantity_bypass.input,
  compare_date.output => l_quantity_bypass.select,
  l_quantity_bypass.output => l_quantity_out,

  instance l_extendedprice_bypass(stream_filter_i<type SQL_decimal_15_2_stream>),
  l_extendedprice_in => l_extendedprice_bypass.input,
  compare_date.output => l_extendedprice_bypass.select,
  l_extendedprice_bypass.output => l_extendedprice_out,

  instance l_discount_bypass(stream_filter_i<type SQL_decimal_15_2_stream>),
  l_discount_in => l_discount_bypass.input,
  compare_date.output => l_discount_bypass.select,
  l_discount_bypass.output => l_discount_out,

  instance l_tax_bypass(stream_filter_i<type SQL_decimal_15_2_stream>),
  l_tax_in => l_tax_bypass.input,
  compare_date.output => l_tax_bypass.select,
  l_tax_bypass.output => l_tax_out,

  instance l_returnflag_bypass(stream_filter_i<type SQL_char1_stream>),
  l_returnflag_in => l_returnflag_bypass.input,
  compare_date.output => l_returnflag_bypass.select,
  l_returnflag_bypass.output => l_returnflag_out,

  instance l_linestatus_bypass(stream_filter_i<type SQL_char1_stream>),
  l_linestatus_in => l_linestatus_bypass.input,
  compare_date.output => l_linestatus_bypass.select,
  l_linestatus_bypass.output => l_linestatus_out,

  instance l_shipdate_bypass(stream_filter_i<type date_stream>),
  l_shipdate_in => l_shipdate_bypass.input,
  compare_date.output => l_shipdate_bypass.select,
  l_shipdate_bypass.output => l_shipdate_out,

  instance l_commitdate_bypass(stream_filter_i<type date_stream>),
  l_commitdate_in => l_commitdate_bypass.input,
  compare_date.output => l_commitdate_bypass.select,
  l_commitdate_bypass.output => l_commitdate_out,

  instance l_receiptdate_bypass(stream_filter_i<type date_stream>),
  l_receiptdate_in => l_receiptdate_bypass.input,
  compare_date.output => l_receiptdate_bypass.select,
  l_receiptdate_bypass.output => l_receiptdate_out,

  instance l_shipinstruct_bypass(stream_filter_i<type SQL_char25_stream>),
  l_shipinstruct_in => l_shipinstruct_bypass.input,
  compare_date.output => l_shipinstruct_bypass.select,
  l_shipinstruct_bypass.output => l_shipinstruct_out,

  instance l_shipmode_bypass(stream_filter_i<type SQL_char10_stream>),
  l_shipmode_in => l_shipmode_bypass.input,
  compare_date.output => l_shipmode_bypass.select,
  l_shipmode_bypass.output => l_shipmode_out,

  instance l_comment_bypass(stream_filter_i<type varchar_stream>),
  l_comment_in => l_comment_bypass.input,
  compare_date.output => l_comment_bypass.select,
  l_comment_bypass.output => l_comment_out,
};

// col: sum_qty, sum_base_price, avg_price
streamlet sum_qty_s {
  l_quantity: SQL_decimal_15_2_stream in,
  l_extendedprice: SQL_decimal_15_2_stream in,
  sum_qty: SQL_decimal_15_2_stream out,
  sum_base_price: SQL_decimal_15_2_stream out,
  avg_price: SQL_decimal_15_2_stream out,
  error: Stream(Bit(1)) out,
};

impl sum_qty_i of sum_qty_s {
  type count_type = streamlet accumulator_s<type SQL_decimal_15_2_stream>.count_type,
  instance accu0(accumulator_i<type SQL_decimal_15_2_stream>),
  instance accu1(accumulator_i<type SQL_decimal_15_2_stream>),

  l_quantity => accu0.input,
  accu0.output => sum_qty,

  l_extendedprice => accu1.input,

  instance avg_price_divider(divider_i<type SQL_decimal_15_2_stream>),
  accu1.output => sum_base_price,
  accu1.output => avg_price_divider.dividend,

  instance converter(converter_i<type count_type, type SQL_decimal_15_2_stream, 1>),
  accu1.count => converter.input[0],
  converter.output[0] => avg_price_divider.divisor,
  avg_price_divider.quotient => avg_price,

  //error
  type error_stream = Stream(Bit(1)),
  instance and(and_i<type error_stream, 2>),
  accu0.overflow => and.input[0] @NoStrictType@,
  accu1.overflow => and.input[1] @NoStrictType@,
  and.output => error @NoStrictType@,
};

// col: sum_disc_price, sum_charge
streamlet sum_disc_price_s {
  l_extendedprice: SQL_decimal_15_2_stream in,
  l_discount: SQL_decimal_15_2_stream in,
  l_tax: SQL_decimal_15_2_stream in,
  sum_disc_price: SQL_decimal_15_2_stream out,
  sum_charge: SQL_decimal_15_2_stream out,

  error: Stream(Bit(1)) out,
};

impl sum_disc_price_i of sum_disc_price_s {
  instance const_decimal_15_value(const_value_generator_i<type SQL_decimal_15_2_stream, 1>),
  instance neg(to_neg_i<type SQL_decimal_15_2_stream>),
  instance adder(adder_i<type SQL_decimal_15_2_stream>),

  //calculate sum_disc_price
  const_decimal_15_value.output => adder.input0,
  l_discount => neg.input,
  neg.output => adder.input1,
  instance multiplier(multiplier_i<type SQL_decimal_15_2_stream>),
  adder.output => multiplier.input0,
  l_extendedprice => multiplier.input1,
  multiplier.output => sum_disc_price, //sum_disc_price

  //calculate sum_charge
  instance multiplier2(multiplier_i<type SQL_decimal_15_2_stream>),
  instance const_decimal_15_value2(const_value_generator_i<type SQL_decimal_15_2_stream, 1>),
  instance adder2(adder_i<type SQL_decimal_15_2_stream>),
  l_tax => adder2.input0,
  const_decimal_15_value2.output => adder2.input1,
  adder2.output => multiplier2.input0,
  multiplier.output => multiplier2.input1,
  multiplier2.output => sum_charge,  //sum_charge

  //error handling
  type error_stream = Stream(Bit(1)),
  instance and(and_i<type error_stream,4>),
  multiplier.overflow => and.input[0] @NoStrictType@,
  adder.overflow => and.input[1] @NoStrictType@,
  adder2.overflow => and.input[2] @NoStrictType@,
  multiplier2.overflow => and.input[3] @NoStrictType@,
  and.output => error @NoStrictType@,
};

// col: avg_qty, avg_disc, count_order
streamlet avg_qty_s {
  l_quantity: SQL_decimal_15_2_stream in,
  l_discount: SQL_decimal_15_2_stream in,
  avg_qty: SQL_decimal_15_2_stream out,
  avg_disc: SQL_decimal_15_2_stream out,
  count_order: Stream(Bit(32)) out,

  error: Stream(Bit(1)) out,
};

impl avg_qty_i of avg_qty_s {
  instance accu0(accumulator_i<type SQL_decimal_15_2_stream>),
  instance divider0(divider_i<type SQL_decimal_15_2_stream>),
  instance accu1(accumulator_i<type SQL_decimal_15_2_stream>),
  instance divider1(divider_i<type SQL_decimal_15_2_stream>),
  instance converter(converter_i<type count_type, type SQL_decimal_15_2_stream, 2>),

  l_quantity => accu0.input,
  accu0.output => divider0.dividend,
  accu0.count => converter.input[1],
  converter.output[1] => divider0.divisor,
  divider0.quotient => avg_qty,

  l_discount => accu1.input,
  accu1.output => divider1.dividend,
  type count_type = streamlet accumulator_s<type SQL_decimal_15_2_stream>.count_type,

  accu1.count => converter.input[0],
  converter.output[0] => divider1.divisor,
  accu1.count => count_order @NoStrictType@,
  divider1.quotient => avg_disc,

  //error
  type error_stream = Stream(Bit(1)),
  instance and(and_i<type error_stream, 2>),
  accu0.overflow => and.input[0] @NoStrictType@,
  accu1.overflow => and.input[1] @NoStrictType@,
  and.output => error @NoStrictType@,
};

//main component
streamlet main_s {
  l_orderkey: key_stream in,
  l_linenumber: key_stream in,

  l_returnflag: SQL_char1_stream out,
  l_linestatus: SQL_char1_stream out,

  sum_qty: SQL_decimal_15_2_stream out,         //part0
  sum_base_price: SQL_decimal_15_2_stream out,  //part0
  sum_disc_price: SQL_decimal_15_2_stream out,  //part1
  sum_charge: SQL_decimal_15_2_stream out,      //part1
  avg_qty: SQL_decimal_15_2_stream out,         //part2
  avg_price: SQL_decimal_15_2_stream out,       //part0
  avg_disc: SQL_decimal_15_2_stream out,        //part2
  count_order: Stream(Bit(32)) out,             //part2

  err: Stream(Bit(1)) out,
};

impl main_i of main_s {
  instance data_src(lineitem_i),
  l_orderkey => data_src.l_orderkey,
  l_linenumber => data_src.l_linenumber,

  instance data_filter(data_filter_i),
  data_src.l_partkey => data_filter.l_partkey_in,
  data_src.l_suppkey => data_filter.l_suppkey_in,
  data_src.l_quantity => data_filter.l_quantity_in,
  data_src.l_extendedprice => data_filter.l_extendedprice_in,
  data_src.l_discount => data_filter.l_discount_in,
  data_src.l_tax => data_filter.l_tax_in,
  data_src.l_returnflag => data_filter.l_returnflag_in,
  data_src.l_linestatus => data_filter.l_linestatus_in,
  data_src.l_shipdate => data_filter.l_shipdate_in,
  data_src.l_commitdate => data_filter.l_commitdate_in,
  data_src.l_receiptdate => data_filter.l_receiptdate_in,
  data_src.l_shipinstruct => data_filter.l_shipinstruct_in,
  data_src.l_shipmode => data_filter.l_shipmode_in,
  data_src.l_comment => data_filter.l_comment_in,

  data_filter.l_returnflag_out => l_returnflag,
  data_filter.l_linestatus_out => l_linestatus,

  //part0
  instance part0(sum_qty_i),
  data_filter.l_quantity_out => part0.l_quantity,
  data_filter.l_extendedprice_out => part0.l_extendedprice,
  part0.sum_qty => sum_qty,
  part0.sum_base_price => sum_base_price,
  part0.avg_price => avg_price,

  //part1
  instance part1(sum_disc_price_i),
  data_filter.l_extendedprice_out => part1.l_extendedprice,
  data_filter.l_discount_out => part1.l_discount,
  data_filter.l_tax_out => part1.l_tax,
  part1.sum_disc_price => sum_disc_price,
  part1.sum_charge => sum_charge,

  //part2
  instance part2(avg_qty_i),
  data_filter.l_quantity_out => part2.l_quantity,
  data_filter.l_discount_out => part2.l_discount,
  part2.avg_qty => avg_qty,
  part2.avg_disc => avg_disc,
  part2.count_order => count_order @NoStrictType@,

  //error
  type error_stream = Stream(Bit(1)),
  instance and(and_i<type error_stream, 3>),
  part0.error => and.input[0] @NoStrictType@,
  part1.error => and.input[1] @NoStrictType@,
  part2.error => and.input[2] @NoStrictType@,
  and.output => err @NoStrictType@,
};
\end{lstlisting}

\begin{figure}[htbp]
    \centering
     \includegraphics[width=0.58\textwidth]{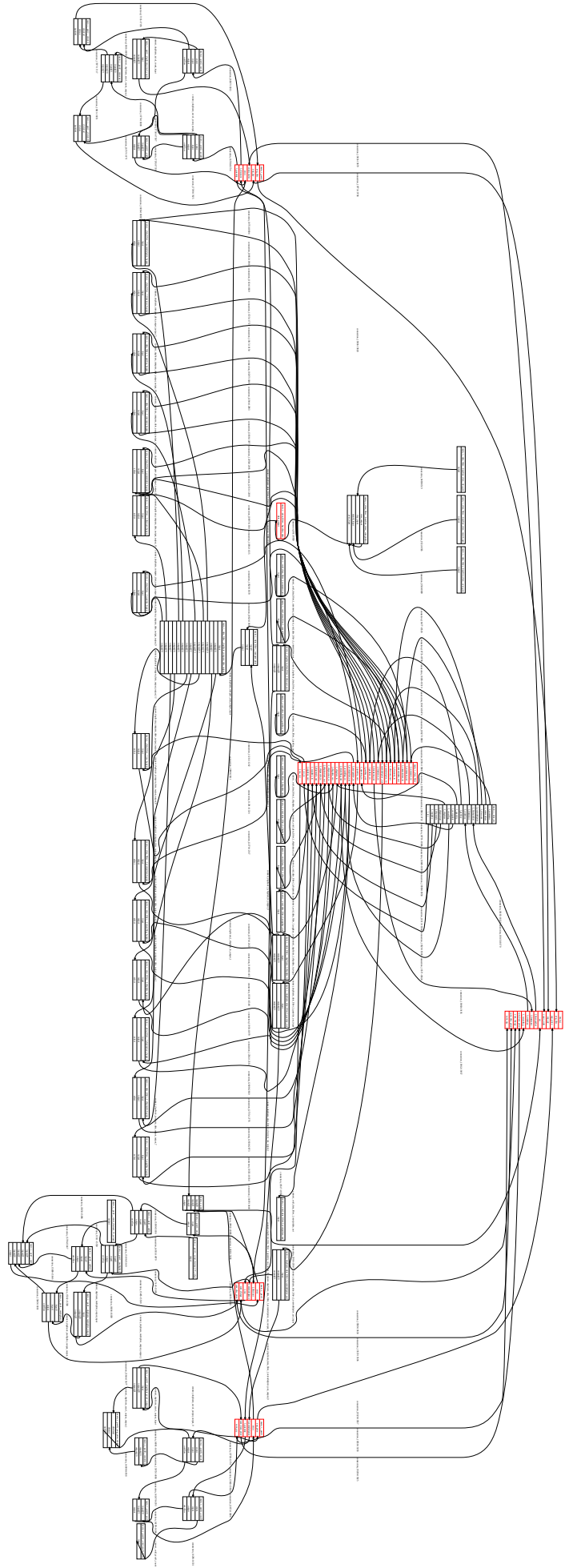}
    \caption{Generated circuit graph for TPC-H query 1}
    \label{fig:circuit_graph_for_TPC_H_query_1}
\end{figure}

Current Tydi-lang simulator supports transforming the code to a circuit representation. Figure \ref{fig:circuit_graph_for_TPC_H_query_1} shows the circuit graph for TPC-H query 1. Each component is represented by a square box. The first line of the box is the name of the component. Other lines represent the names of ports. The directed arrows indicates connections. The text on the connection indicates the connection name (specified in code). Red boxes mean wrapper components. The ports on wrapper components have two arrows: one flows in and one flows out. In traditional hardware synthesis tools, components represented by red boxes are same as components that are clickable (click to see the internal structure). 

All TPC-H queries mentioned in this thesis have their own circuit graph, available in the TPC-H folders of the following link: \href{https://github.com/twoentartian/tydi-lang/tree/main/CookBook}{https://github.com/twoentartian/tydi-lang/tree/main/CookBook}. 

%% file: 07-conclusion.tex
\chapter{Conclusion}

This thesis presents a new language (Tydi-lang) based on Tydi-spec to allow developers more effectively design streaming hardware. This new language also introduces the template concept to typed hardware, which raises the level of abstraction, saving design efforts for developers and enabling the possibility of translating software domain languages to hardware description languages. The syntax, grammar details and many sample codes are provided in this thesis for future Tydi-lang users and future Tydi-lang compiler developers. The structure of the compiler and possible optimization methods are also discussed in this paper.

Along with the high-level language, we also present the blueprint of Tydi-lang simulator, a verification and simulation tool, to show how Tydi-lang works with low-level languages and improve the efficiency of high-level design. We implement the Tydi-lang compiler prototype with its standard library and use several SQL query cases to demonstrate the new design flow and its effectiveness.

I would like to give the answers to the questions in the problem statement (Section \ref{subsection:problem_statement}).

\begin{itemize}
    \item What is the essential language syntax to describe typed streaming hardware based on the Tydi-spec?\newline
    Answer: Section \ref{section:tydi_language_specifiction} shows the Tydi-lang syntax. The syntax to define logical types, streamlets (including inner ports), and implementations (including inner instance and connections) is essential. The other syntax, such as defining variables, templates, "for" and "if" blocks, and assertions, is designed to reduce the design effort and reduce the possibility of making mistakes.
    
    \item How to minimize the design effort for Tydi-lang users?\newline
    Answer: Raise the level of abstraction by using variables and templates to describe hardware. Provide compiler-level sugaring and standard template library for developers.
    
    \item Rust is a relatively new language, its unique immutable/mutable reference system requires more design effort on the memory structure. How can we address the memory challenges specific to designing a compiler in Rust?\newline 
    Answer: Use read/write locks on each node of the code structure tree to get the immutable reference when needed. This immutable reference also allows multithreading optimization mentioned in Section \ref{subsubsection:multi_thread_evaluation}. Properly solving this issue can avoid many other future problems in designing the compiler.
    
    \item What kind of abstraction method should the compiler provide to facilitate designing typed streaming hardware?\newline
    Answer: The abstraction methods include template and generative "if"/"for" syntax. The template supports accepting seven types of arguments: integer, float point numbers, string, boolean, clockdomain, logical type, and components. These template arguments are enough to describe hardware in most cases. The "if" and "for" syntax can generate parallel hardware circuits according to variable values.
    
    \item Hardware simulation and verification is an important phase in design flow, how to assess hardware simulation and verification in the context of a Tydi-spec based toolchain.\newline 
    Answer: First of all, Tydi-lang provides a high-level design rule check. Developers can find mistakes such as logical type mistakes quickly. The Tydi-lang simulation syntax mentioned in Section \ref{subsection:tydi_simulation_syntax} enables the possibility to perform high-level simulations to predict data sequence or analyze the streaming bottleneck. Meanwhile, the Tydi-lang simulator can generate low-level testbenches from the simulation syntax to allow cooperation between high-level designers and low-level designers.
    
    \item How to enable the cooperation between the new language and other existing HDLs and tools?\newline
    Answer: Besides the hardware verification and simulation cooperation mentioned above, Tydi-lang also provides mechanisms to deliver the high-level description to low-level HDLs, such as the documentation and logical type system. There are many ways to perform language-level cooperation. For example, the RTL generator of the Tydi-lang standard library can be written in CHISEL, and developing a new backend from Tydi-IR to CHISEL ensures high portability.
    
\end{itemize}

There are many possible future works for Tydi-lang, and I record them here for further development:

\begin{itemize}
    \item Support all functions in the Tydi-lang simulator. For example, we can include support for the stack-based virtual machine, the simulation recorder, the testbench generator.
    \item Implement the RTL generator for the Tydi-lang standard library.
    \item The current cross-package reference feature in Tydi-lang is not well tested because of the missing support on the backend side.
    \item Improving Fletcher to generate Tydi interfaces.
    \item Support generating CHISEL and/or integrate Tydi support for CHISEL. 
\end{itemize}


%% file: 08-appendix.tex
\chapter{Appendix}

\section{Known issues in the Tydi-lang compiler}
There are some potential issues in the current implementation of the Tydi-lang compiler. I put them in the appendix for future developers' reference.

\subsection{Wrong precedence for unary operator}
\label{appendix:wrong_precedence_for_unary_operator}
The following code snippet illustrates a precedence error for unary operators (\texttt{!} and \texttt{-}) during evaluating mathematical expressions.

\begin{lstlisting}[language=c++]
package test;

const i1 = -1+2; //evaluated result: i1 = -3, correct: 1
const i2 = i1 + 5;
type bit = Bit(i2);
type stream = Stream(bit);
\end{lstlisting}

The cause is that the Tydi-lang compiler evaluate \texttt{1+2} first and finally evaluate the \texttt{-(1+2)}. This error may be fixed by modifying the PEST grammar files, but I tried several times and didn't get a working version. 

Traditional language workbench systems, such as Spoofax \cite{spoofax}, support left-recursive parsing, while the PEST parser does not support this feature yet because it causes great performance loss. The missing of left-recursive parsing results in invalid grammar rules like \texttt{Exp = Exp + Exp} because evaluating the second \texttt{Exp} will recursively use this rule.

Due to this limitation, the Tydi-lang compiler separate the semantic \texttt{Exp} to two terms, \texttt{Exp} and \texttt{Term}, with the following PEST grammar:

\begin{lstlisting}[language=c++]
//available in tydi_lang_syntax.pest: 59~65
Term = { ( "(" ~ Exp ~ ")" ) |
            ... | UnaryExp}
Exp = { Term ~ (InfixOp  ~ Term)* }
\end{lstlisting}

In theory, the \texttt{UnaryExp} should have the least precedence because no other expression is starting with \texttt{-} or \texttt{!}. Though in practice, it does not work correctly. The missing left-recursive in PEST parser makes this issue more complicated because I was forced to separate the semantic \texttt{Exp} into two parts. I recommend future developers start by reconsidering the definition of \texttt{Term} and \texttt{Exp} to fix this bug.

\subsection{Duplicated identifier issue in for/if expansion}
\label{subsubsection:Duplicated_identifier_issue_in_for_if_expansion}
In Tydi-lang, users can define connections in for and if scopes. The name of these connections will be the connection name appended with the scope name. The scope name is different in each for/if expansion, so the connections name will be different after expansion. However, defining instances in a for scope will result in errors because the name does not change in each expansion.

Defining instances inside "for"/"if" scope is useful in many cases. For example, in TPCH benchmark - query 19, there are three similar query statement blocks but with different arguments. Tydi-lang developers can use a "for" syntax to define the three query structure whose arguments are defined in an array. It is impossible in the current Tydi-lang compiler because we cannot define instances in "for"/"if" scope.

Thus, I propose an alternative identifier syntax subject to variable values. In Tydi-lang, the traditional identifiers can only contain digits, alphabet char, and underscore. The alternative identifier should have the following PEST grammar rule.

\begin{lstlisting}[language=c++]
ID = @{ ID_BLOCK_LIST ~ (ID_INVALID_CHAR ~ ( ASCII_ALPHA | "_" )) ~ (ID_INVALID_CHAR ~ ( ASCII_ALPHA | ASCII_DIGIT | "_" ))*  ~ !(ASCII_ALPHA | ASCII_DIGIT | "_") }

VAR_IN_ALTERNATIVE_ID = @{"{{" ~ ID ~ "}}"}
ALTERNATIVE_ID = { ID_BLOCK_LIST ~ (ID_INVALID_CHAR ~ ( ASCII_ALPHA | "_" )) ~ (ID_INVALID_CHAR ~ ( ASCII_ALPHA | ASCII_DIGIT | "_" | VAR_IN_VAR_ID))*  ~ !(ASCII_ALPHA | ASCII_DIGIT | "_") }
\end{lstlisting}

The \texttt{VAR\_IN\_ALTERNATIVE\_ID} is an identifier to a basic variable and its value will be evaluated construct the \texttt{ALTERNATIVE\_ID}. In a "for" scope, users can use the state variable in the "for" statement to give different names to instances to avoid the duplicated identifier problem after code expansion.

An example of the use of the alternative identifier syntax.
\begin{lstlisting}[language=c++]
package main;

type bit8_stream = Stream(Bit(8), d = 5, t = 2.5);

streamlet data_bypass<data: str> {         
  input: bit8_stream in,
  output: bit8_stream out,
};
impl impl_data_bypass<data: str> of data_bypass<data> {
  input => output,
};

const channel = 4;
streamlet data_bypass_channel {
  inputs: bit8_stream [channel] in `"10kHz",
  outputs: bit8_stream [channel] out `"10kHz",
};

const use_data_bypass2 = true;
const data = {"Monday", "Tuesday", "Wednesday", "Thursday"};

impl impl_data_bypass_channel of data_bypass_channel {
  //the external scope contains 4 instances: bypass_0,bypass_1,bypass_2,bypass_3, each of them will have different template arguments.
  for i in (0=1=>channel) {
    instance bypass_{{i}}(impl_data_bypass<data[i]>),  //when i == 1, the bypass_{{i}} will be evaluated to bypass_1
    bypass_{{i}}.output => outputs[i],        //bypass_{{i}} => bypass_1
    inputs[i] => bypass_{{i}}.input,
  }
};
\end{lstlisting}

\section{Proposals about future work}

\begin{table}[H]
\centering
\caption{Proposed builtin functions in Tydi-lang}
\label{table:proposed_builtin_functions}
\resizebox{\textwidth}{!}{
\begin{tabular}{|l|l|l|}
\hline
\multicolumn{1}{|c|}{function identifier(arg list) -\textgreater output} & \multicolumn{1}{c|}{explanation} & \multicolumn{1}{c|}{example} \\ \hline
\begin{tabular}[c]{@{}l@{}}get\_child\_names(\{logical\_type\}) \\ -\textgreater Array\textless{}string\textgreater{}\end{tabular} & \begin{tabular}[c]{@{}l@{}}return the child names if the logical\\ type is group or union.\end{tabular} & \begin{tabular}[c]{@{}l@{}}get\_child\_names(rgb)\\ -\textgreater{}\{"r","g","b"\}\end{tabular} \\ \hline
\begin{tabular}[c]{@{}l@{}}get\_child(\{logical\_type\}, \{child\_name\})\\ -\textgreater{}logical\_type\end{tabular} & \begin{tabular}[c]{@{}l@{}}return the child logical type with \\ given child name\end{tabular} & \begin{tabular}[c]{@{}l@{}}get\_child(rgb,"r")\\ -\textgreater{}Bit(8)\end{tabular} \\ \hline
\begin{tabular}[c]{@{}l@{}}type\_of\_logical\_type(\{logical\_type\})\\ -\textgreater{}String\end{tabular} & \begin{tabular}[c]{@{}l@{}}return "group" if the logical type is a \\ group type, "union" for logical union,\\ "stream" for logical stream, "bit" for \\ logical bit.\end{tabular} & \begin{tabular}[c]{@{}l@{}}type\_of\_logical\_type(rgb)\\ -\textgreater{}"group"\end{tabular} \\ \hline
\end{tabular}
}
\end{table}

\begin{table}[H]
\centering
\caption{Proposed builtin functions in Tydi simulator}
\label{table:proposed_builtin_functions_tydi_simulator}
\begin{tabular}{|l|l|l|}
\hline
\multicolumn{1}{|c|}{function identifier(arg list) -\textgreater{}output} & \multicolumn{1}{c|}{explanation} & \multicolumn{1}{c|}{example} \\ \hline
receive(port\_name) -\textgreater bool & \begin{tabular}[c]{@{}l@{}}returns "true" if the data packet is \\ available on the port. Equivalent \\ to getting a handshaking signal on\\  the hardware side.\end{tabular} & receive(data\_in\_0) \\ \hline
\begin{tabular}[c]{@{}l@{}}read(port\_name) -\textgreater \\ \{composite data representation\}\end{tabular} & \begin{tabular}[c]{@{}l@{}}return the received composite data\\ from this port.\end{tabular} & read(data\_in\_0) \\ \hline
\begin{tabular}[c]{@{}l@{}}send(port\_name, \\ \{composite data representation\});\end{tabular} & \begin{tabular}[c]{@{}l@{}}send certain composite data packet\\ via this port.\end{tabular} & \begin{tabular}[c]{@{}l@{}}send(data\_out\_0, Group(\\ a=0x11110000,\\ b=0x11110000));\end{tabular} \\ \hline
delay\_cycle(int, \{Frequency\}) & \begin{tabular}[c]{@{}l@{}}delay for certain cycles of a \\ frequency. The frequency must be \\ one of the clockdomain values \\ available on the component.\end{tabular} & delay\_cycle(5, 100MHz); \\ \hline
delay(time) & \begin{tabular}[c]{@{}l@{}}delay for a certain physical time. \\ The physical time should be \\ achievable with the available \\ frequencies on that component, \\ otherwise, the Tydi simulator \\ should throw an error.\end{tabular} & delay\_cycle(1us); \\ \hline
ack(port\_name) & \begin{tabular}[c]{@{}l@{}}add 1 to the acknowledge counter\\ for the port. The physical \\ acknowledge signal will be sent \\ when the value becomes large \\ enough.\end{tabular} & ack(data\_in\_0); \\ \hline
\end{tabular}
\end{table}

%% file: main.bbl
\begin{thebibliography}{20}
\providecommand{\natexlab}[1]{#1}
\providecommand{\url}[1]{\texttt{#1}}
\expandafter\ifx\csname urlstyle\endcsname\relax
  \providecommand{\doi}[1]{doi: #1}\else
  \providecommand{\doi}{doi: \begingroup \urlstyle{rm}\Url}\fi

\bibitem[abs tudelft(2022)]{tydi-prototype}
abs tudelft.
\newblock Tydi-an open specification and tools for complex data structures over
  hardware streams., 2022.
\newblock URL \url{https://github.com/abs-tudelft/tydi}.

\bibitem[Apache(2022{\natexlab{a}})]{arrow}
Apache.
\newblock Apache arrow, 2022{\natexlab{a}}.
\newblock URL \url{https://arrow.apache.org/}.

\bibitem[Apache(2022{\natexlab{b}})]{spark}
Apache.
\newblock Apache spark - unified engine for large-scale data analytics,
  2022{\natexlab{b}}.
\newblock URL \url{https://spark.apache.org/}.

\bibitem[Bachrach et~al.(2012)Bachrach, Vo, Richards, Lee, Waterman,
  Avižienis, Wawrzynek, and Asanović]{chisel}
Jonathan Bachrach, Huy Vo, Brian Richards, Yunsup Lee, Andrew Waterman, Rimas
  Avižienis, John Wawrzynek, and Krste Asanović.
\newblock Chisel: Constructing hardware in a scala embedded language.
\newblock In \emph{DAC Design Automation Conference 2012}, pages 1212--1221,
  2012.
\newblock \doi{10.1145/2228360.2228584}.

\bibitem[Graphviz(2022)]{dot-lang}
Graphviz.
\newblock Dot language, 2022.
\newblock URL \url{https://www.graphviz.org/doc/info/lang.html}.

\bibitem[Hoozemans et~al.(2021)Hoozemans, Peltenburg, Nonnemacher, Hadnagy,
  Al-Ars, and Hofstee]{bigdata_fpgas}
Joost Hoozemans, Johan Peltenburg, Fabian Nonnemacher, Akos Hadnagy, Zaid
  Al-Ars, and H.~Peter Hofstee.
\newblock Fpga acceleration for big data analytics: Challenges and
  opportunities.
\newblock \emph{IEEE Circuits and Systems Magazine}, 21\penalty0 (2):\penalty0
  30--47, 2021.
\newblock \doi{10.1109/MCAS.2021.3071608}.

\bibitem[Houtgast et~al.(2017)Houtgast, Sima, and Al-Ars]{genomics_fpgas}
Ernst Houtgast, Vlad-Mihai Sima, and Zaid Al-Ars.
\newblock High performance streaming smith-waterman implementation with
  implicit synchronization on intel fpga using opencl.
\newblock In \emph{2017 IEEE 17th International Conference on Bioinformatics
  and Bioengineering (BIBE)}, pages 492--496, 2017.
\newblock \doi{10.1109/BIBE.2017.000-6}.

\bibitem[Kats and Visser(2010)]{spoofax}
Lennart~C.L. Kats and Eelco Visser.
\newblock The spoofax language workbench: Rules for declarative specification
  of languages and ides.
\newblock In \emph{Proceedings of the ACM International Conference on Object
  Oriented Programming Systems Languages and Applications}, OOPSLA '10, page
  444–463, New York, NY, USA, 2010. Association for Computing Machinery.
\newblock ISBN 9781450302036.
\newblock \doi{10.1145/1869459.1869497}.
\newblock URL
  \url{https://doi-org.tudelft.idm.oclc.org/10.1145/1869459.1869497}.

\bibitem[Microsoft(2022)]{dot-lang-vscode-ext}
Microsoft.
\newblock Graphviz (dot) language support for visual studio code, 2022.
\newblock URL
  \url{https://marketplace.visualstudio.com/items?itemName=joaompinto.vscode-graphviz}.

\bibitem[Nvidia(2022)]{cuda}
Nvidia.
\newblock Cuda toolkit - free tools and trainings for developers, 2022.
\newblock URL \url{https://developer.nvidia.com/cuda-toolkit}.

\bibitem[Peltenburg et~al.(2019{\natexlab{a}})Peltenburg, van Straten, Brobbel,
  Hofstee, and Al-Ars]{fletcher_internals}
Johan Peltenburg, Jeroen van Straten, Matthijs Brobbel, H.~Peter Hofstee, and
  Zaid Al-Ars.
\newblock Supporting columnar in-memory formats on fpga: The hardware design of
  fletcher for apache arrow.
\newblock In Christian Hochberger, Brent Nelson, Andreas Koch, Roger Woods, and
  Pedro Diniz, editors, \emph{Applied Reconfigurable Computing}, pages 32--47,
  Cham, 2019{\natexlab{a}}. Springer International Publishing.
\newblock ISBN 978-3-030-17227-5.

\bibitem[Peltenburg et~al.(2019{\natexlab{b}})Peltenburg, van Straten,
  Wijtemans, van Leeuwen, Al-Ars, and Hofstee]{fletcher}
Johan Peltenburg, Jeroen van Straten, Lars Wijtemans, Lars van Leeuwen, Zaid
  Al-Ars, and Peter Hofstee.
\newblock Fletcher: A framework to efficiently integrate fpga accelerators with
  apache arrow.
\newblock In \emph{2019 29th International Conference on Field Programmable
  Logic and Applications (FPL)}, pages 270--277, 2019{\natexlab{b}}.
\newblock \doi{10.1109/FPL.2019.00051}.

\bibitem[Peltenburg et~al.(2020{\natexlab{a}})Peltenburg, van Leeuwen,
  Hoozemans, Fang, Al-Ars, and Hofstee]{parquet_fpgas}
Johan Peltenburg, Lars~T.J. van Leeuwen, Joost Hoozemans, Jian Fang, Zaid
  Al-Ars, and H.~Peter Hofstee.
\newblock Battling the cpu bottleneck in apache parquet to arrow conversion
  using fpga.
\newblock In \emph{2020 International Conference on Field-Programmable
  Technology (ICFPT)}, pages 281--286, 2020{\natexlab{a}}.
\newblock \doi{10.1109/ICFPT51103.2020.00048}.

\bibitem[Peltenburg et~al.(2020{\natexlab{b}})Peltenburg, Van~Straten, Brobbel,
  Al-Ars, and Hofstee]{tydi-spec}
Johan Peltenburg, Jeroen Van~Straten, Matthijs Brobbel, Zaid Al-Ars, and
  H.~Peter Hofstee.
\newblock Tydi: An open specification for complex data structures over hardware
  streams.
\newblock \emph{IEEE Micro}, 40\penalty0 (4):\penalty0 120--130,
  2020{\natexlab{b}}.
\newblock \doi{10.1109/MM.2020.2996373}.

\bibitem[Peltenburg et~al.(2021)Peltenburg, Hadnagy, Brobbel, Morrow, and
  Al-Ars]{tens_of_gigabytes}
Johan Peltenburg, Ákos Hadnagy, Matthijs Brobbel, Robert Morrow, and Zaid
  Al-Ars.
\newblock Tens of gigabytes per second json-to-arrow conversion with fpga
  accelerators.
\newblock In \emph{2021 International Conference on Field-Programmable
  Technology (ICFPT)}, pages 1--9, 2021.
\newblock \doi{10.1109/ICFPT52863.2021.9609833}.

\bibitem[pest parser(2022)]{pest}
pest parser.
\newblock pest, the elegant parser, 2022.
\newblock URL \url{https://pest.rs/}.

\bibitem[Reukers(2022)]{tydi-backend}
Matthijs Reukers.
\newblock Wip/playground vhdl back-end for (yet to be defined) tydi
  intermediate representation., 2022.
\newblock URL \url{https://github.com/matthijsr/til-vhdl}.

\bibitem[rust lang.org(2022)]{rust-lifetime}
rust lang.org.
\newblock Lifetimes, rust by examples, 2022.
\newblock URL
  \url{https://doc.rust-lang.org/rust-by-example/scope/lifetime.html}.

\bibitem[salsa rs(2022)]{rust-salsa}
salsa rs.
\newblock Salsa-a generic framework for on-demand, incrementalized
  computation., 2022.
\newblock URL \url{https://github.com/salsa-rs/salsa}.

\bibitem[Solutions(2022)]{tpch}
Hotea Solutions.
\newblock Tpch homepage, 2022.
\newblock URL \url{https://www.tpc.org/tpch/}.

\end{thebibliography}
